\documentclass[twocolumn,twocolappendix]{aastex7}

\usepackage{amsmath}
\shorttitle{EP250302a}
\shortauthors{O'Connor et al.}
\submitjournal{ApJ}

\usepackage{siunitx}
\usepackage{booktabs}
\usepackage{longtable}
\usepackage{caption}

\definecolor{blazeorange}{rgb}{1.0, 0.4, 0.0}
\definecolor{seagreen}{rgb}{0.18, 0.55, 0.34}
\definecolor{darkgreen}{rgb}{0.08, 0.45, 0.2}
\definecolor{rufous}{rgb}{0.66, 0.11, 0.03}
\definecolor{royalfuchsia}{rgb}{0.79, 0.17, 0.57}
\definecolor{scarlet}{rgb}{1.0, 0.13, 0.0}
\definecolor{royalpurple}{rgb}{0.47, 0.32, 0.66}

\begin{document}

\title{Discovery of a Supernova Following the Einstein Probe Transient EP250302a at $z=1.131$
}

\correspondingauthor{Brendan O'Connor}
\author[0000-0002-9700-0036]{Brendan O'Connor}
    \altaffiliation{McWilliams Fellow}
    \affiliation{McWilliams Center for Cosmology and Astrophysics, Department of Physics, Carnegie Mellon University, Pittsburgh, PA 15213, USA}
    \email[show]{boconno2@andrew.cmu.edu}  

\author[0009-0001-0574-2332
]{Malte Busmann}
    \altaffiliation{Recipient of a Wübben Stiftung Wissenschaft Student Grant}
    \affiliation{University Observatory, Faculty of Physics, Ludwig-Maximilians-Universität München, Scheinerstr. 1, 81679 Munich, Germany}
    \email{m.busmann@physik.lmu.de}
    \affiliation{Excellence Cluster ORIGINS, Boltzmannstr. 2, 85748 Garching, Germany}
	\email{daniel.gruen@lmu.de}

\author[0009-0006-7990-0547]{James Freeburn}
    \affiliation{University of North Carolina at Chapel Hill, 120 E. Cameron Ave., Chapel Hill, NC 27514, USA}
    \affiliation{Sydney Institute for Astronomy, School of Physics, University of Sydney, Sydney, NSW 2006, Australia}
    \affiliation{ARC Centre of Excellence for Gravitational Wave Discovery (OzGrav), Hawthorn, Victoria, 3122, Australia}
    \email{jamesfreeburn54@gmail.com}

\author[0000-0001-7201-1938]{Lei Hu}
    \affiliation{McWilliams Center for Cosmology and Astrophysics, Department of Physics, Carnegie Mellon University, Pittsburgh, PA 15213, USA}
    \email{leihu@andrew.cmu.edu}

\author[0000-0002-0786-7307]{Noel Klingler}
    \affiliation{Center for Space Sciences and Technology, University of Maryland, Baltimore County, Baltimore, MD 21250, USA}
    \affiliation{Astrophysics Science Division, NASA Goddard Space Flight Center, 8800 Greenbelt Rd, Greenbelt, MD 20771, USA}
    \affiliation{Center for Research and Exploration in Space Science and Technology, NASA/GSFC, Greenbelt, Maryland 20771, USA}
    \email{noelklin@umbc.edu}

\author[0000-0003-3270-7644]{Daniel Gruen}
	\affiliation{University Observatory, Faculty of Physics, Ludwig-Maximilians-Universität München, Scheinerstr. 1, 81679 Munich, Germany}
	\affiliation{Excellence Cluster ORIGINS, Boltzmannstr. 2, 85748 Garching, Germany}
	\email{daniel.gruen@lmu.de}
    

\author[0000-0002-8977-1498]{Igor Andreoni}
    \affiliation{University of North Carolina at Chapel Hill, 120 E. Cameron Ave., Chapel Hill, NC 27514, USA}
    \email{igor.andreoni@unc.edu}

\author[0009-0008-2754-1946]{Julius Gassert}
    \affiliation{University Observatory, Faculty of Physics, Ludwig-Maximilians-Universität München, Scheinerstr. 1, 81679 Munich, Germany}
    \affiliation{McWilliams Center for Cosmology and Astrophysics, Department of Physics, Carnegie Mellon University, Pittsburgh, PA 15213, USA}
    \email{julius.gassert@campus.lmu.de}

\author[0000-0002-9364-5419]{Xander J. Hall}
	\affiliation{McWilliams Center for Cosmology and Astrophysics, Department of Physics, Carnegie Mellon University, Pittsburgh, PA 15213, USA}
	\email{xjh@andrew.cmu.edu}

\author[0000-0002-6011-0530]{Antonella Palmese}
	\affiliation{McWilliams Center for Cosmology and Astrophysics, Department of Physics, Carnegie Mellon University, Pittsburgh, PA 15213, USA}
	\email{palmese@cmu.edu}


\begin{abstract}

We present a multi-wavelength analysis of the Einstein Probe (EP) fast X-ray transient (FXT) EP250302a located at redshift $z=1.131$. Despite its luminous prompt X-ray emission, the event was not detected in gamma-rays. Multi-wavelength follow-up identified a bright optical and X-ray source that displayed rapid chromatic flaring before returning to the standard decay of a gamma-ray burst afterglow. We interpret the chromatic flare as either due to a refreshed shock caused by a discrete shell collision or as reverse shock emission.  Using the early optical data, we place constraints on the Lorentz factor of the outflow, requiring an ultrarelativistic jet with $\Gamma_0>25$. 
We additionally obtained deep late-time imaging with the Gemini North Telescope that reveals the presence of an optical excess at $20-30$ d post-explosion. We interpret this as supernova (SN) emission and find good agreement with the canonical broad-lined Ic SN 1998bw with a flux-scaling factor of $k_\textrm{98bw}>0.3$. This adds to the growing evidence that the majority of EP FXTs are associated with the deaths of massive stars.

\end{abstract}

\keywords{\uat{X-ray transient sources}{1852} --- \uat{Gamma-ray bursts}{629} --- \uat{Relativistic jets}{1390} --- \uat{Core-collapse supernovae}{304} --- \uat{Type Ic supernovae}{1730} }


\section{Introduction}

Fast X-ray transients (FXTs) are luminous, rapidly fading soft X-ray flashes with durations of seconds to hours \citep[e.g.,][]{Jonker2013,Glennie2015,Bauer2017,Alp2020}. Their physical origins have remained uncertain because most were identified serendipitously in archival X-ray observations and therefore lacked prompt multi-wavelength follow-up \citep[e.g.,][]{Jonker2013,Glennie2015,Bauer2017,Quirola2022,Quirola2023}. The launch of the Einstein Probe (EP; \citealt{Yuan2025}) has changed this situation by surveying the soft X-ray sky with a wide-field lobster-eye imager. Although the majority of EP FXTs lack detected gamma-ray counterparts \citep[e.g.,][]{Yin2024,Liu2024,Jiang2025,Zhang2024,OConnor2025,Yadav2025}, their redshift distribution \citep{OConnor2025EP-z}, afterglow-like behavior \citep[e.g.,][]{Levan2024,Gillanders2024,vandalen2024,srivastav2024,Busmann2025}, and consistency with the Amati relation \citep{Sun2024,Liu2024,Li2025,OConnor2025EP-z} indicate that most extragalactic EP FXTs are closely related to long-duration gamma-ray bursts (GRBs) and arise from similar massive star progenitors \citep{Woosley1993,MacFadyen1999}. 

Despite the numerous circumstantial arguments, the clearest evidence for this interpretation is the emerging connection between EP FXTs and stripped-envelope supernovae, hereafter FXT-SNe. For reference, long GRBs are securely linked to broad-lined Type Ic supernovae \citep{Woosley2006,Hjorth2012sn,Cano2017}, beginning with GRB 980425/SN 1998bw \citep{Galama1998,Patat2001} and extending through dozens of discoveries over the following three decades \citep[referred to as GRB-SNe, see, e.g.,][]{Hjorth2003,Stanek2003,Campana2006,Soderberg2006grb060218,Starling2011,DElia2018,Izzo2019}. In the past two years since launch, numerous Ic-BL supernovae have now been uncovered following EP FXTs over a broad range of prompt and afterglow properties \citep{Sun2024,vandalen2024,Li2025,Rastinejad2025EP,Srinivasaragavan2025EP0108a,Srinivasaragavan2026,Quirola2026sn,vanHoof2026,Cotter2026,OConnor2026,Yuan2026,Rastinejad2026,MartinCarrillo2026,Chen2026}, supporting their relation to similar  progenitors to those producing long GRBs and GRB-SNe \citep{OConnor2025EP-z}.

EP250302a was detected by the EP Wide-field X-ray Telescope (WXT) on 2025-03-02 at 15:36:04 UT \citep[hereafter $T_0$;][]{2025GCN.39556....1D,Fu2026}. A bright optical afterglow was rapidly detected starting at $\sim$\,$460$ s after the EP trigger \citep{2025GCN.39555....1W,2025GCN.39550....1Z}. Shortly after, multiple teams reported an early optical rebrightening \citep{2025GCN.39569....1K,2025GCN.39565....1P} that was simultaneous with an X-ray flare \citep{2025GCN.39556....1D,Fu2026}. A redshift of $z$\,$=$\,$1.131$ was acquired based on absorption lines seen in optical spectra of the afterglow  \citep{0302a-vlt-redshift,2025GCN.39574....1Y,OConnor2025EP-z,Fu2026}. The X-ray and optical rebrightening was interpreted by \citet{Fu2026} as a violent collision of two relativistic shells \citep{ZhangMeszaros2002}. They suggest that this is the first instance of a discrete refreshed shock in an FXT. However, it must be noted that previous rebrightenings observed in EP transients have been interpreted similarly \citep[e.g.,][]{srivastav2024,Busmann2025}. 

In this work, we present the results of a comprehensive multi-wavelength observing campaign of EP250302a using the Fraunhofer Telescope Wendelstein, Gemini North Telescope, \textit{Neil Gehrels Swift Observatory}, and the \textit{Chandra X-ray Observatory}. Using this dataset, we identify a late-time optical excess that is consistent with expectations for a Ic-BL SN at $z$\,$=$\,$1.131$. We conclude that despite the lack of prompt gamma-rays, EP250302a is a GRB-like massive star explosion.  

Throughout the manuscript we adopt a standard $\Lambda$CDM cosmology \citep{Planck2020} with $H_0$\,$=$\,$67.4$ km s$^{-1}$ Mpc$^{-1}$, $\Omega_\textrm{m}$\,$=$\,$0.315$, and $\Omega_\Lambda$\,$=$\,$0.685$.
We also adopt the convention for the flux density is parametrized by $F_\nu$\,$\propto$\,$t^{-\alpha}\nu^{-\beta}$ where $\alpha$ is the powerlaw temporal index and $\beta$ is the powerlaw spectral index.
All upper limits are reported at the $3\sigma$ level and all magnitudes are in the AB system.

\section{Observations}
\label{sec:obs}

\begin{figure*}
    \centering
\includegraphics[width=2.1\columnwidth]{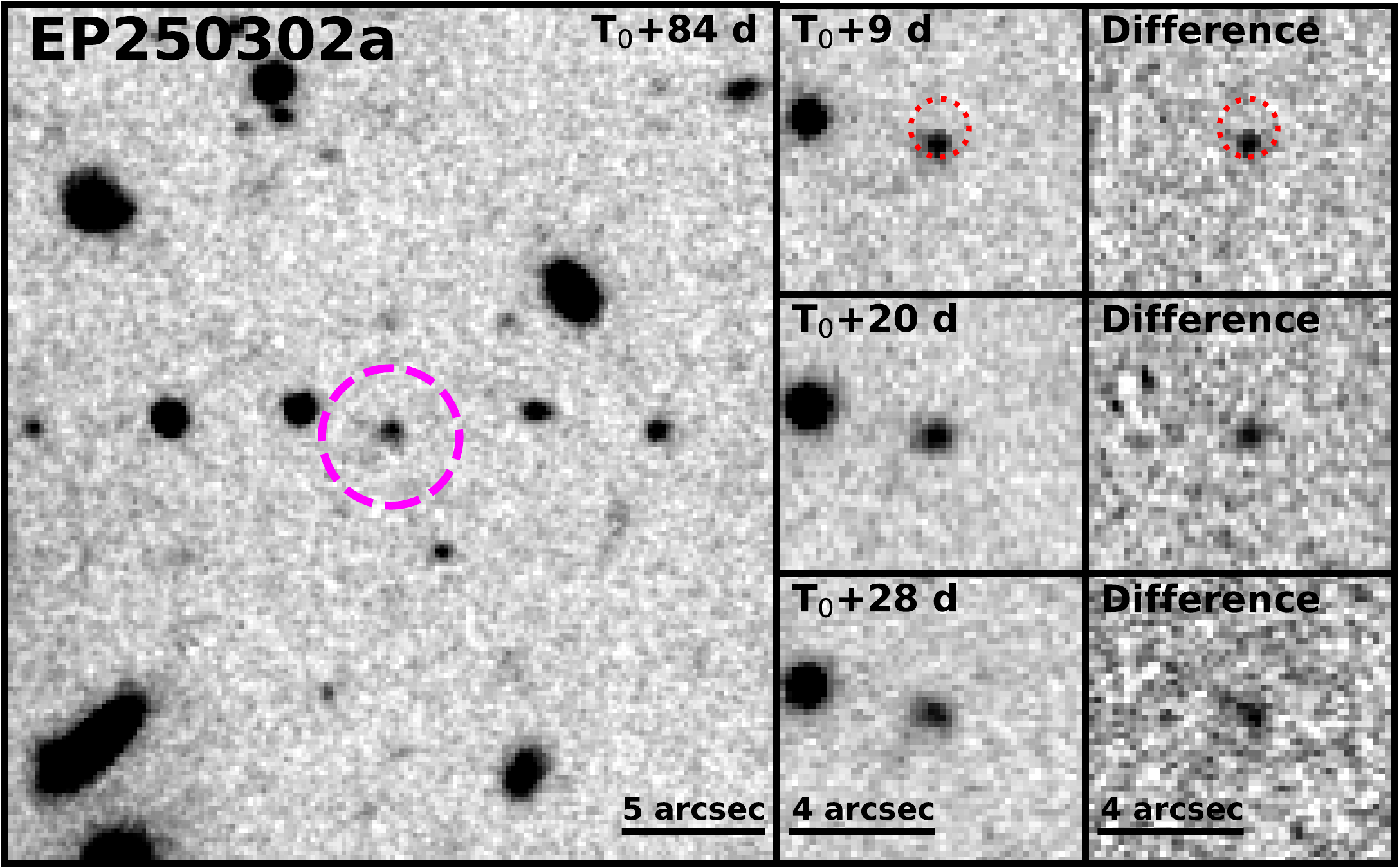}
    \caption{\textbf{Left:} Finding chart of EP250302a using our deep Gemini GMOS-N $i$-band reference image ($i$\,$>$\,$26.3$ mag) obtained at $T_0$\,$+$\,$84$ d. The magenta circle shows the \textit{Swift}/XRT localization (\S \ref{sec:XRT}). \textbf{Right:} The six small panels display the the three initial Gemini epochs ($i$-band) at $T_0$\,$+$\,$9$ d, $20$ d, and $28$ d (observer frame) after the EP trigger. The difference images are shown on the far right panels and present a clear residual source. The red circle in the top panels shows the subarcsecond \textit{Chandra} X-ray localization (\S \ref{sec:chandra}). In all panels, North is up and East is to the left.
    }
    \label{fig:FC}
\end{figure*}

\subsection{Fraunhofer Telescope Wendelstein (FTW)}
\label{sec:FTW}

We observed the optical and near-infrared (OIR) counterpart of EP250302a with the Three Channel Imager (3KK; \citealt{2016SPIE.9908E..44L}) mounted on the 2.1-m Fraunhofer Telescope at Wendelstein Observatory (FTW; \citealt{2014SPIE.9145E..2DH}). Observations began on 2025-03-02 at 19:28:31 UT, corresponding to $\sim$3.87 hr to 5.4 d after the EP trigger \citep{2025GCN.39551....1B}. All observations across each epoch simultaneously obtained imaging data in the $riJ$ filters using the three-channel 3KK camera. Additional late-time observations were obtained between 2025-04-28 and 2025-05-01 during dark time for a total exposure of 12.6 hr in the $riJ$ filters. The mid-time of the late-time observation is 2025-04-30 at 05:32:09 UT, corresponding to 58.58 d. The complete log of observations is tabulated in Table \ref{tab: observationsPhot}.

The FTW data were reduced using a custom pipeline \citep{2002A&A...381.1095G} to perform bias and dark subtraction, flat-fielding, and cosmic ray rejection. For more details, specifically with regard to the near-infrared data reduction, see \citet{Busmann2025}. In the deep late-time epoch at $T_0+58$ d, we detect a weak source in the $riJ$ bands. We performed difference imaging using the Saccadic Fast Fourier Transform (\texttt{SFFT}) software\footnote{\url{https://github.com/thomasvrussell/sfft}} \citep{Hu2022} with a deep Gemini $i$-band image (\S \ref{sec:gemini}) obtained at $T_0+84$ d and verified there was no residual source in FTW $i$-band at $T_0+58$ d to depth $i>25.1$ mag. We consider the multi-band detections at $T_0+58$ d to be the host galaxy.

Due to the early epoch of the FTW detections, the source photometry is not significantly impacted by the host galaxy contribution (see also \S \ref{sec:gemini}). We verified this by performing difference imaging with respect to our late-time images in all bands. 
We performed point spread function (PSF) photometry on the difference images using \texttt{Photutils} \citep{Bradley2024} with AB magnitude zeropoints calibrated to the PS1 \citep{Chambers2016} and 2MASS \citep{Skrutskie2006} catalogs for optical and near-infrared data, respectively. NIR magnitudes have been converted from the Vega to AB magnitude system using standard offsets \citep{Skrutskie2006}. The photometry is reported in Table \ref{tab: observationsPhot}.

\subsection{Gemini North Telescope}
\label{sec:gemini}

We performed multi-epoch optical imaging with the 8.1-m Gemini North Telescope located on Mauna Kea using the Gemini Multi-Object Spectrograph (GMOS-N; \citealt{Hook2004}). Observations were obtained through a Director's Discretionary Time (DDT) request (GN-2025A-DD-104; PI: O'Connor) across four epochs in $i$-band on 2025-03-11, 2025-03-22, 2025-03-30, and 2025-05-25 (see Table \ref{tab: observationsPhot}). In our initial epoch, the background was impacted by close proximity to a very bright moon (requiring a higher exposure time; $30\times90$ s), but all other epochs ($15\times150$ s) were obtained during dark time at large distances from the moon. The data were reduced using the \texttt{Dragons V3.2.0} software package \citep{Labrie2019,Labrie2023}. 

A clear, high signal-to-noise source is detected at the position of EP250302a (\S \ref{sec:FTW}) in each epoch (Figure \ref{fig:FC}). In the final epoch, the source appears point-like with half-light radius $\sim$\,$0.4\arcsec$ ($\sim$\,$3$ kpc). We consider the source detection in the final epoch to be the compact host galaxy of EP250302a after fading of the transient. This is supported by the offset between the transient and host galaxy (see below). By performing aperture photometry on each of the epochs, we observe a clear decline by $\sim$0.7 mag between 2025-03-11 and 2025-05-25. In the final epoch we measure a brightness of $i$\,$=$\,$25.2\pm0.1$ mag for the host galaxy. The $3\sigma$ depth of this image is $i$\,$>$\,$26.3$ mag.

Using our final epoch as a template, we performed difference imaging using the Saccadic Fast Fourier Transform (\texttt{SFFT}) software\footnote{\url{https://github.com/thomasvrussell/sfft}} \citep{Hu2022}. We detect a residual source in our difference imaging in the initial three epochs on 2025-03-11, 2025-03-22, and 2025-03-30, see Figure \ref{fig:FC}. Using the difference image from the first observation (highest signal-to-noise; Figure \ref{fig:FC}) on 2025-03-11, the offset between the host galaxy and transient is found to be $0.24\pm0.03\arcsec$, corresponding to $2.0\pm0.2$ kpc at $z$\,$=$\,$1.131$ \citep{OConnor2025EP-z}. 
We performed PSF photometry using \texttt{Photutils} \citep{Bradley2024} with photometric zeropoints calibrated to the PS1 catalog \citep{Chambers2016}. The photometry is reported in Table \ref{tab: observationsPhot}.

\subsection{\textit{Neil Gehrels Swift Observatory}}

\subsubsection{X-ray Telescope (XRT)}
\label{sec:XRT}

The \textit{Neil Gehrels Swift Observatory} (hereafter \textit{Swift}; \citealt{Gehrels2004}) rapidly responded to the EP/WXT trigger through an automatic Priority 0 (P0) Target of Opportunity (ToO) observation with settled observations starting at $T_0+1$ hr  \citep{2025GCN.39557....1P}. A clear, fading X-ray source was localized to an enhanced position of RA, DEC (J2000) = $11^{h}18^m 03^{s}.62$, $+33^\circ 35\arcmin 09.0\arcsec$ with uncertainty of $2.5\arcsec$ (90\% CL), consistent with the optical localization of the transient \citep{2025GCN.39557....1P}. We submitted a further ToO request for multiple additional observations in subsequent days. In total, the X-ray Telescope \citep[XRT;][]{Burrows2005} acquired data in Photon Counting (PC) mode with 13.5 ks obtained over 7 epochs (Target ID: 19585). We retrieved the automatic \textit{Swift}/XRT data processing and analysis from the Leicester Site\footnote{\url{https://www.swift.ac.uk/EP/EP_FIELD00023/Source1/}}.  

The X-ray spectrum is only well constrained in the initial high count rate ($0.1$\,$-$\,$1.0$ cts s$^{-1}$) observations at $T_0$\,$<$\,$10^{4}$ s. We applied the Swift-XRT Build Products tool\footnote{\url{https://www.swift.ac.uk/user_objects/index.php}} to extract and model the spectrum. The time-averaged X-ray spectrum ($0.3$\,$-$\,$10$ keV) can be modeled by an absorbed power-law with photon index $\Gamma_\textrm{X}$\,$=$\,$1.83\pm0.20$ with a Galactic absorber $N_\textrm{H,gal}$\,$=$\,$2.47\times10^{20}$ cm$^{-2}$ \citep{Willingale2013} and intrinsic absorber $N_\textrm{H,int}$\,$=$\,$(1.8^{+2.7}_{-1.8})\times10^{21}$ cm$^{-2}$ at $z$\,$=$\,$1.131$ \citep{OConnor2025EP-z,Fu2026}. We note that the intrinsic column density is potentially consistent with a negligible contribution, and largely unconstrained by the data. We tested this by modeling the data with only an unconstrained Galactic absorber ($z$\,$=$\,$0$), yielding $N_\textrm{H}$\,$=$\,$4^{+5}_{-4}\times10^{20}$ cm$^{-2}$ and $\Gamma_\textrm{X}$\,$=$\,$1.83\pm0.24$. This is consistent with the estimated Galactic value along the line-of-sight of $N_\textrm{H,gal}$\,$=$\,$2.47\times10^{20}$ cm$^{-2}$ \citep{Willingale2013}. Therefore, we conclude that the data do not require additional absorption (see also \citealt{Fu2026}). 

Using the best fit spectrum, the count rate to unabsorbed flux energy conversion factor is $4.11\times10^{-11}$ erg cm$^{-2}$ cts$^{-1}$, which we apply uniformly to the extracted count rate lightcurve. A log of X-ray observations is tabulated in \ref{tab: observationsXray}. The X-ray lightcurve is shown in Figure \ref{fig:xray}.

\subsubsection{Ultra-Violet Optical Telescope (UVOT)}
\label{sec:uvot}

The \textit{Swift} Ultra-Violet Optical Telescope (UVOT; \citealt{Roming2005}) obtained settled observations beginning at $T_0+1$ hr \citep{2025GCN.39570....1S}. A fading optical source was discovered consistent with the XRT enhanced position \citep{2025GCN.39570....1S}. The initial P0 observations were carried out in $u$-band, and we requested the use of this filter be continued in our subsequent ToO observations. In total, 13.5 ks of $u$-band imaging was taken. We broke each observation into its individual snapshots, yielding 17 total snapshots in this dataset. In each snapshot, we performed aperture photometry at the source position with the \texttt{uvotsource} task calibrated with the standard AB magnitude zeropoints \citep{Breeveld2011}. We identify that in 11 out of 17 snapshots the source position unfortunately falls on a small-scale sensitivity (SSS) patch\footnote{\url{https://www.swift.ac.uk/analysis/uvot/sss.php}} \citep{Breeveld2010} on the UVOT detector. We have applied the default screening value of \texttt{ssstype=LOW}. When the source falls on a SSS patch, \texttt{uvotsource} returns 99 for the measured magnitudes. We reject these snapshots and do not report photometry for them as the SSS patches can have reduced sensitivity by up to 39\%, leading to inaccurate photometry. In any case, we find that the source position is only impacted by the SSS patch after it has already faded, and we detect the fading source in the first two snapshots as first reported by  \citet{2025GCN.39570....1S}. 
The UVOT photometry is reported in Table \ref{tab: observationsPhot}.

\subsection{\textit{Chandra X-ray Observatory}}
\label{sec:chandra}

The \textit{Chandra X-ray Observatory} (CXO) observed EP250302a through a DDT request (ObsID: 30840; PI: O'Connor) starting on 2025-03-11 at 14:04:47 UT ($T_0+8.9$ d) for 20 ks with ACIS-S. The \textit{Chandra} data were retrieved from the 
\textit{Chandra} Data Archive (CDA)\footnote{\url{https://cda.harvard.edu/chaser/}}. We re-processed the data using the \texttt{CIAO v4.17.0} data reduction package with \texttt{CALDB v4.11.6}. The data were filtered to the $0.5$\,$-$\,$8$ keV energy range. We detect a source ($\sim$\,$20$ cts) consistent with the optical localization of EP250302a at RA, DEC (J2000) = $11^{h}18^m 03^{s}.6$, $+33^\circ 35\arcmin 09.6\arcsec$ with an uncertainty of $\sim$\,$0.8\arcsec$ \citep{2025GCN.39698....1O}. Assuming the absorbed power-law spectrum found by \textit{Swift}/XRT (see \S \ref{sec:XRT}, we computed the unabsorbed flux in the $0.3$\,$-$\,$10$ keV using the \texttt{srcflux} task and a $1.5\arcsec$ circular extraction region. We derive $F_\textrm{X}$\,$=$\,$(2.4^{+0.6}_{-0.5})\times 10^{-14}$ erg cm$^{-2}$ s$^{-1}$. This is further tabulated in Table \ref{tab: observationsXray}. This data were also analyzed by \citet{Fu2026} who derive a consistent flux.

\begin{figure}
    \centering
    \includegraphics[width=\linewidth]{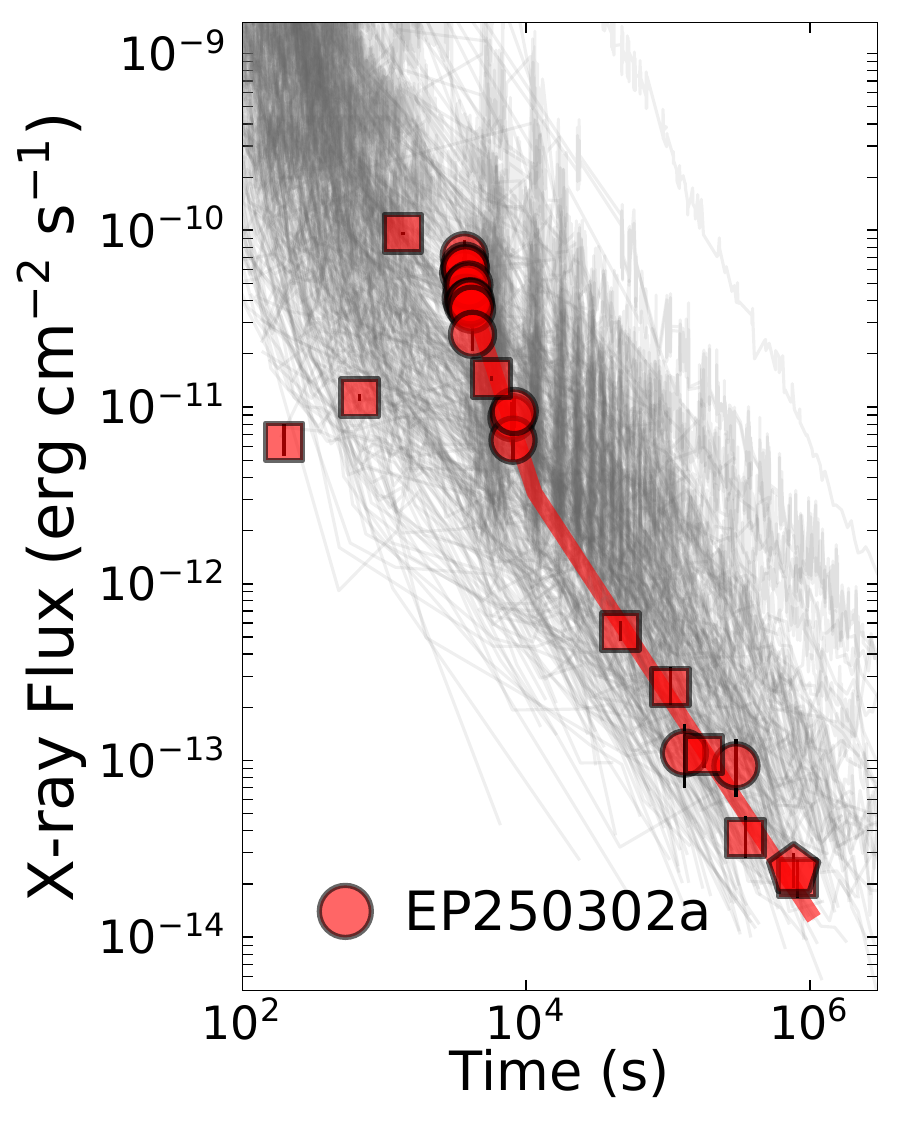}
    \caption{Observed X-ray lightcurves ($0.3$\,$-$\,$10$ keV) of \textit{Swift} long duration ($>$\,$2$ s) GRBs versus EP250302a (red circles). The solid red line depicts the best-fit lightcurve to EP250302a. We have included data from \textit{Swift} (circles) and \textit{Chandra} (pentagon) analyzed in this work and EP/FXT (squares) data presented by \citet{Fu2026}.  }
    \label{fig:xray}
\end{figure}

\section{Results}
\label{sec:results}

\subsection{Temporal Properties}
\label{sec:temporalfits}

\subsubsection{Optical to Near-infrared Lightcurve}

Our OIR data of EP250302a were obtained between 3.87 hr and 84 d with FTW and Gemini-North. Multiple teams reported an early optical rebrightening \citep{2025GCN.39569....1K,2025GCN.39565....1P}, see also \citet{Fu2026} for a recent analysis of a comprehensive dataset. Here we consider only post-flare data and discuss the flare briefly in Appendix \ref{sec:appendix-flare}. We modeled the late-time decay after $0.1$ d using the FTW ($riJ$ bands) data out to 3 days. The fit was performed across these multiple bands simultaneously. The decay is well described by a single powerlaw with slope $\alpha_\textrm{OIR,late}$\,$=$\,$1.08\pm0.05$. We revisit this in \S \ref{sec:snmodel}.

\subsubsection{X-ray Lightcurve}

The X-ray lightcurve was likewise found to display a sharp rebrightening based on early EP/FXT follow-up observations \citep{2025GCN.39556....1D,Fu2026}.  X-ray rebrightenings and flares are common in the early afterglows of \textit{Swift} GRBs \citep[e.g.,][]{Burrows2005flare,Nousek2006}. 
The \textit{Swift}/XRT observations starting at $T_0$\,$+$\,$3.6$ ks caught the end of the rebrightening during its decaying phase (see also \citealt{Fu2026}).

A temporal fit to the X-ray data yields a temporal break at $t_\textrm{X,b}$\,$=$\,$(1.1^{+0.9}_{-0.5})\times10^{4}$ s (observer frame). The initial slope is steep $\alpha_{X,1}$\,$=$\,$2.44^{+0.32}_{-0.24}$ due to \textit{Swift}/XRT observations starting at the end of the observed rebrightening during its period of steep decay. 
The post-break slope shows a more typical X-ray afterglow decay of $\alpha_{X,2}$\,$=$\,$1.2^{+0.1}_{-0.1}$ at the $1\sigma$ level. The late-time decay slope $\alpha_{X,2}$ agrees with the decay observed at OIR wavelengths. The best-fit lightcurve is shown in Figure \ref{fig:xray}.

\begin{figure*}
    \centering
    \includegraphics[width=\linewidth]{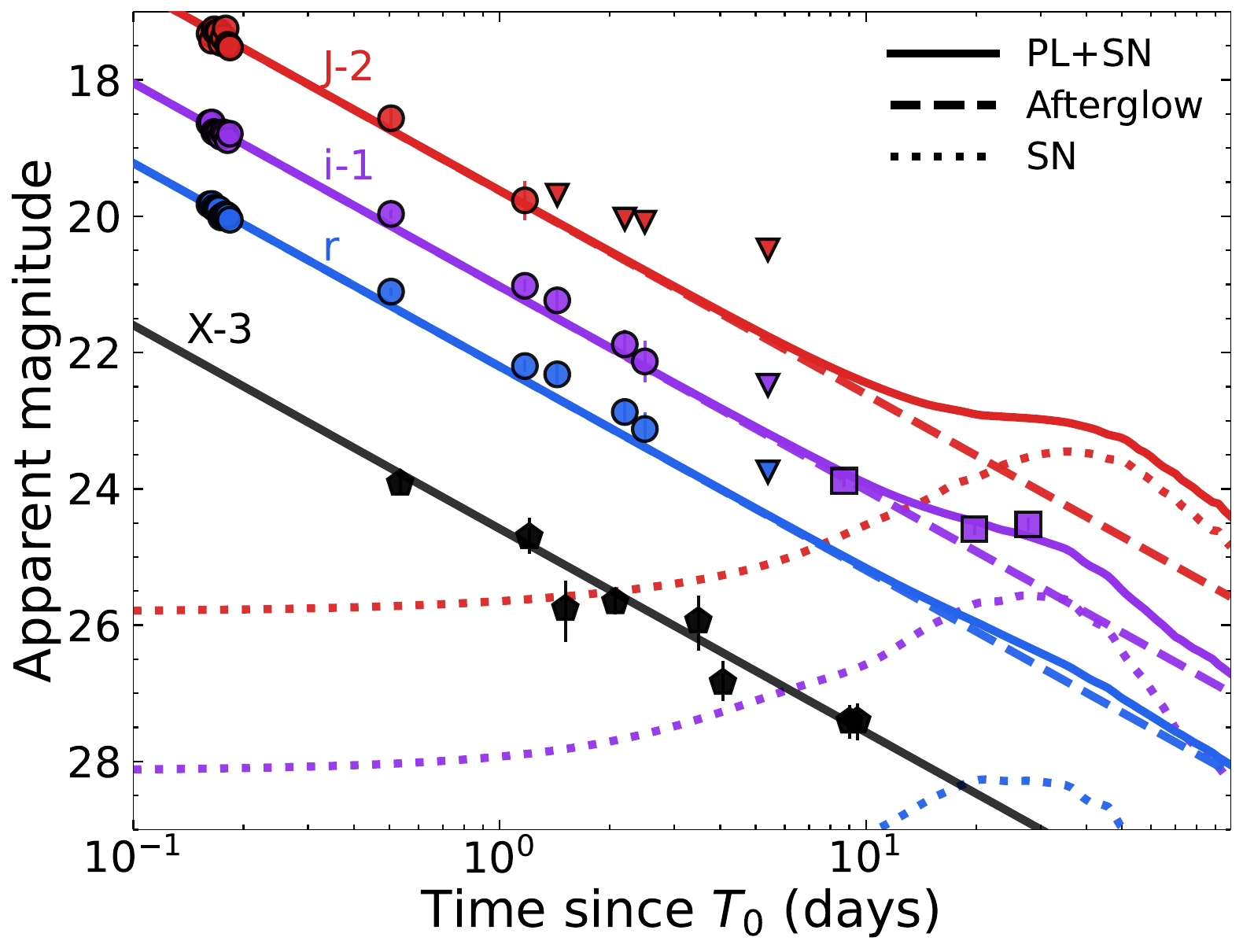}
    \caption{Best-fit models to the multi-wavelength ($riJ$ + X-ray) lightcurve of EP250302a in the observer-frame. FTW data are shown as circles, Gemini data as squares, and X-ray data (This work and \citealt{Fu2026}) are shown as pentagons. Upper limits from FTW are shown as downward triangles. The total model (solid lines) consists of an afterglow component (dashed lines) and the addition of a supernova (dotted lines), modeled using SN 1998bw with a flux scaling factor $k_\textrm{98bw}$\,$=$\,$0.40$ (see \S \ref{sec:snmodel}). 
    }
    \label{fig:snmodel}
\end{figure*}

\subsection{Spectral Energy Distribution Properties}
\label{sec:SEDs}

We used the multi-band OIR data of EP250302a to constrain its spectral energy distribution (SED) as a function of time. All OIR data were corrected for the Galactic extinction $E(B-V)$\,$=$\,$0.022$ mag \citep{Schlafly2011} along the line-of-sight. We make use of simultaneous $riJ$ band observations from FTW at $\sim$\,$0.16$\,$-$\,$0.18$ d, $0.50$ d, and $1.17$ d. We additionally include a simultaneous epoch of $ugr$ data at $\sim$\,$0.09$ d obtained by \textit{Swift}/UVOT and NUTTelA-TAO \citep{2025GCN.39569....1K}. We modeled these data across these multiple epochs simultaneously with a powerlaw $F_\nu$\,$\propto$\,$\nu^{-\beta_\textrm{OIR}}$. We derive a best-fit constant spectral index of $\beta_\textrm{OIR}$\,$=$\,$0.76\pm0.04$. The best-fit has a $\chi^2/\textrm{dof}$\,$=$\,$1.2$ for 51 degrees of freedom.

Using the FTW data, we find no evidence for an evolution of this spectral index with consistent values between $0.16$ d ($3.87$ hr) and $1.17$ d. This is based on an analysis of individual epochs of $riJ$ data performed in addition to our fit including all epochs. 
This OIR spectral index is consistent with the photon indices inferred from X-ray observations with $\Gamma_\textrm{X}$\,$=$\,$\beta_\textrm{X}+1$\,$=$\,$1.83\pm0.20$ from \textit{Swift}/XRT. However, we must point out that the X-ray photon indices are derived from X-ray data obtained during the rebrightening phase, and the late-time X-ray photon index might differ (though this is not always the case; see Figure 10 of \citealt{Falcone2007}). 

Based on the agreement with the X-ray photon index during the flare and the late-time OIR spectral index, we suggest the X-ray and OIR data likely lie on the same power-law segment (see Figure \ref{fig:snmodel}). We therefore find no strong evidence for requiring an additional intrinsic dust component, especially provided the consistency of the $riJ$ derived spectral index with the $ugr$ derived spectral index. This is especially relevant when considering that the observed $i$-band is approximately the rest frame $u$-band at this redshift ($z$\,$=$\,$1.131$; \citealt{OConnor2025EP-z,Fu2026}), and the observed $u$-band lies deep in the ultraviolet regime where any intrinsic extinction (i.e., in the host galaxy at $z$\,$=$\,$1.131$) would suppress the emission. This is also consistent with the lack of additional intrinsic absorption required by the X-ray spectra (see \S \ref{sec:XRT} for details).

\begin{figure}
    \centering
    \includegraphics[width=\linewidth]{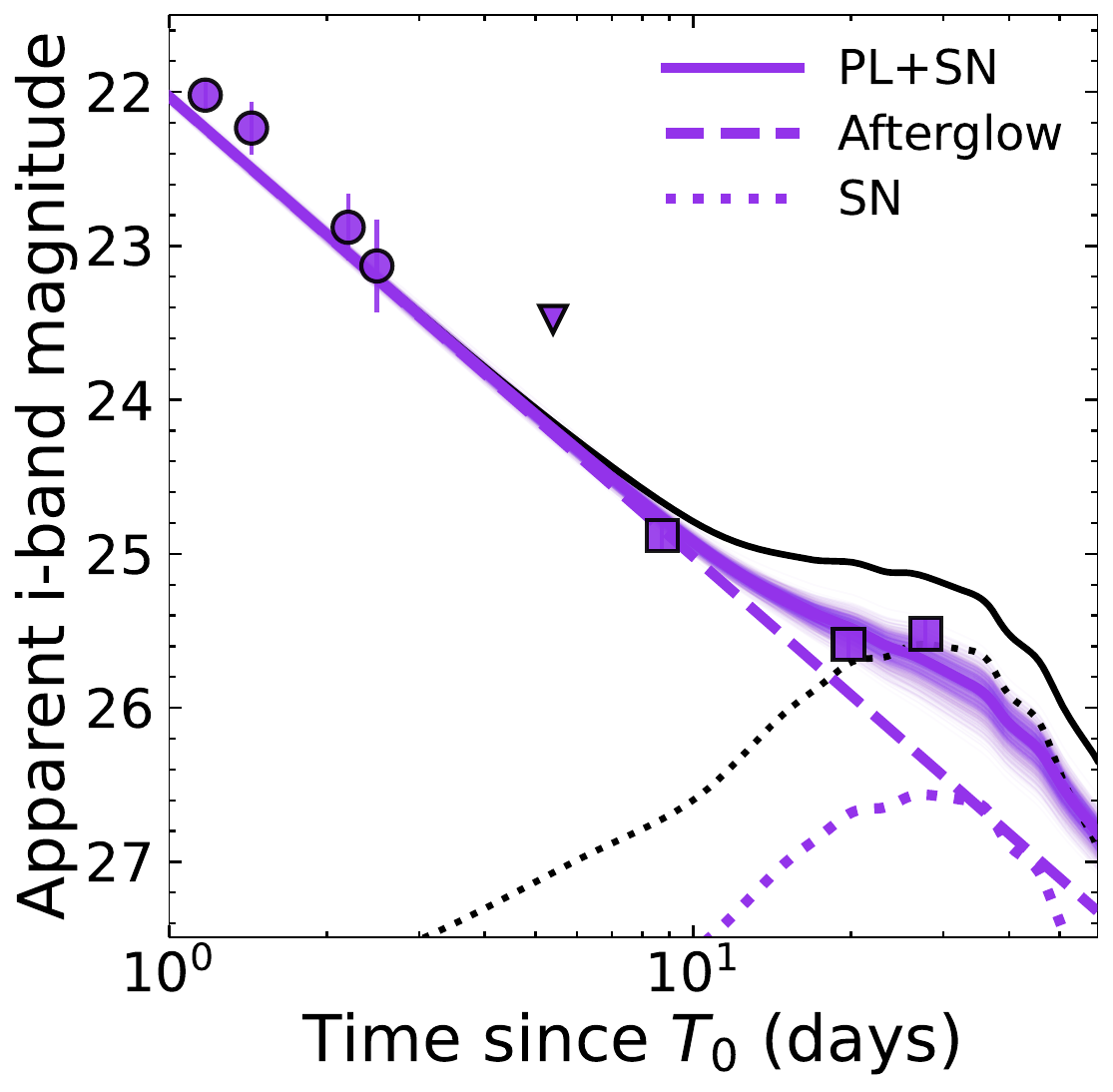}
    \caption{Same as Figure \ref{fig:snmodel} but for only the $i$-band data and the PL+SN model (purple) which assumes there is no jet-break. The individual afterglow (dashed) and SN (dotted) components are shown along with 1000 PL+SN model iterations randomly sampled from the posterior. The SN component is produced using a SN 1998bw template \citep{Vincenzi2019}. The best-fit purple curves are shown for $k_\textrm{98bw}$\,$=$\,$0.40$. A representative case of  $k_\textrm{98bw}$\,$=$\,$1.0$ is shown by  the black curves.
    }
    \label{fig:snmodel-zoom}
\end{figure}

\subsection{Multi-wavelength Afterglow Properties}
\label{sec:afterglowfit}


Here, we explore the consistency between the observed temporal and spectral indices and GRB afterglow closure relations \citep[e.g.,][]{Granot2002}. The standard fireball model of GRBs \citep{Meszaros1997,Sari1998,Wijers1999} predicts broadband non-thermal emission, referred to as the afterglow. This afterglow emission is produced by an external shock caused by the deceleration of the relativistic fireball (the jet) into the surrounding lower density environment with density profile $\rho_\textrm{ext}(R)$\,$=$\,$A\, R^{-k}$. A distribution of relativistic electrons is accelerated in the shock to form an initial powerlaw distribution of electron energies $N(E)$\,$\propto$\,$E^{-p}$ that then cool through synchrotron radiation, producing a characteristic broadband behavior that evolves with time \citep[e.g.,][]{Granot2002}. The observed emission, and its temporal and spectral properties, depends on the observing frequency and the ordering of the characteristic synchrotron frequencies: the self-absorption frequency $\nu_\textrm{a}$, the injection frequency $\nu_\textrm{m}$, and the cooling frequency $\nu_\textrm{c}$.

We focus on the late-time decay ($>$\,$0.1$ d) after the end of the optical and X-ray rebrightening \citep{Fu2026}, and check the consistency with forward shock (FS) afterglow emission. In a uniform density environment, typical of the interstellar medium (ISM), for observing frequencies between the injection frequency and the cooling frequency in the regime $\nu_\textrm{m}$\,$<$\,$\nu$\,$<$\,$\nu_\textrm{c}$, the temporal and spectral indices are provided by $\alpha$\,$=$\,$3(p-1)/4$ and $\beta$\,$=$\,$(p-1)/2$ \citep[e.g.,][]{Granot2002}. This can be re-written as $p$\,$=$\,$2\beta+1$\,$=$\,$4\alpha/3+1$, which yields $\alpha$\,$=$\,$3\beta/2$ in order to be consistent with emission from this regime. Applying our best-fit values for $\alpha$ and $\beta$ based on the OIR data, we find $p$\,$=$\,$2\beta+1$\,$=$\,$2.52\pm0.08$ and $p$\,$=$\,$4\alpha/3+1$\,$=$\,$2.44\pm0.04$. These predictions are consistent at the $1\sigma$ level. Additionally, the prediction for the temporal decay from the spectral index is also consistent at the $1\sigma$ level with $\alpha$\,$=$\,$3\beta/2$\,$=$\,$1.14\pm0.06$. As we do not find consistency with any other synchrotron segment for an ISM environment, or any parametrization for a stellar wind environment ($k$\,$=$\,$2$), we consider that EP250302a occurred within an ISM-like environment with X-ray to OIR emission in the regime $\nu_\textrm{m}$\,$<$\,$\nu$\,$<$\,$\nu_\textrm{c}$ (at least between 0.1 and 2 d).

\subsection{Supernova Modeling}
\label{sec:snmodel}

Following \citet{OConnor2026}, we performed a simultaneous multi-wavelength optical to near-infrared (OIR) fit to the lightcurve of EP250302a in the $riJ$ filters using models supplied in the \texttt{redback} software package \citep{redback}. Posterior sampling was performed with \texttt{Bilby} \citep{bilby} using the nested sampler \texttt{dynesty} \citep{dynesty}. We empirically modeled the lightcurve with the combination of a afterglow component, modeled as a smoothly broken powerlaw, plus the contribution of a supernova using the \texttt{sncosmo} \citep{sncosmo} \texttt{v19-1998bw} template \citep{Vincenzi2019} which is based on observations of SN 1998bw \citep{Galama1998,Stathakis2000,Patat2001}. This model is defined by a flux scaling factor $k_\textrm{98bw}$, where a value of $k_\textrm{98bw}$\,$=$\,$1$ is the exact flux scale of SN 1998bw with rest-frame absolute magnitudes $M_B$\,$=$\,$-18.9$ and $M_V$\,$=$\,$-19.3$ mag \citep{Galama1998}. 

For this purpose, we include only our FTW $riJ$ data obtained between 0.16 to 5.4 d after the EP trigger and the later time Gemini $i$-band data obtained at 8.7, 19.8, and 27.8 d. At $z$\,$=$\,$1.131$ this corresponds to roughly rest-frame $u$-band. We additionally include the X-ray lightcurve (Figure \ref{fig:xray}) to aid in constraining the temporal and spectral indices. We used two different afterglow setups. The first assumes that there is no jet-break \citep{Rhoads1999} such that the afterglow is a single powerlaw (PL), whereas the second fit allows for a jet-break such that the afterglow is parametrized as a smoothly broken powerlaw (SBPL). We have assumed the jet to be laterally spreading \citep{SariPiranHalpern1999}. Both models use the standard ISM closure relations between $p$, $\alpha$, and $\beta$ \citep{Granot2002}. These are outlined in \S \ref{sec:afterglowfit}.

The best-fit PL+SN lightcurve is shown in Figure \ref{fig:snmodel} and the results of our lightcurve modeling are displayed in Figure \ref{fig:corner-sn} in Appendix \ref{sec:appendix-sn-model-corner}. Overall we identify that $p$\,$=$\,$2.6$ provides a good description of the data, and that an equally good fit can be obtained for both a jet-break as early as $4$ d and as late as after the end of all the multi-wavelength data. 
The reduced chi-squared is unity for both model fits. In either case, there is a late-time flux excess in the rest-frame $u$-band that occurs on a similar timescale of GRB-SN \citep{Hjorth2012sn,Cano2017}. Even under the assumption that there is no jet-break in this time frame, the afterglow model underpredicts the $i$-band flux at $30$ d by 0.8 magnitudes (see Figure \ref{fig:snmodel-zoom}). By modeling this excess as a supernova, we derive a range of flux-scaling values with respect to SN 1998bw between $k_\textrm{98bw}$\,$=$\,$0.3$\,$-$\,$1.0$, see the corner plot displayed in Figure \ref{fig:corner-sn} in Appendix \ref{sec:appendix-sn-model-corner}. This is consistent with the range of values found for GRB associated Ic-BL SNe \citep{Hjorth2012sn,Cano2017}. As a jet-break is expected, but not robustly observed for this event, we consider $k_\textrm{98bw}$\,$\gtrsim$\,$0.3$ a lower bound to the supernova brightness (see Figure \ref{fig:snmodel-zoom}).

We note that any residual supernova emission in our template epoch at 84 days would have decreased our measured brightness in the difference images (Figure \ref{fig:FC}). However, assuming $k_\textrm{98bw}$\,$=$\,$1.0$, the supernova emission in $i$-band at 84 days would be $i$\,$=$\,$27.9$ AB mag (see Figure \ref{fig:snmodel-zoom}) and it would be 1 magnitude fainter if $k_\textrm{98bw}$\,$=$\,$0.4$. Therefore, we find that any residual transient flux at late times would have only a marginal impact (e.g., between $5$\,$-$\,$10\%$) on our inferred brightness at $20$\,$-$\,$30$ d.

\section{Discussion}
\label{sec:discussion}

\subsection{Relation to GRBs}

Despite the lack of prompt gamma-ray detection (see Appendix \ref{sec:gammalimits}), the other properties of EP250302a, including the prompt X-ray emission duration of 42 s and isotropic-equivalent energy of $E_\textrm{WXT}=(5.0\pm1.0)\times10^{50}$ erg in the rest-frame $0.5$\,$-$\,$4$ keV \citep{Fu2026}, are consistent with the typical expectations for a GRB \citep{Kouveliotou1993}. The follow-up X-ray and optical temporal and spectral properties (\S \ref{sec:temporalfits} and \ref{sec:SEDs}) match well expectations for a GRB afterglow (\S \ref{sec:afterglowfit}) after $0.1$ d. Figure \ref{fig:xray} shows that the observed X-ray lightcurve of EP250302a has good agreement with \textit{Swift} GRBs \citep{Nousek2006}.

The rare property of EP250302a (in addition to the lack of gamma-rays), that is certainly not shown by all GRB afterglows, is the early, steep rebrightening at $<$\,$0.1$ d 
(Figure \ref{fig:rbandfit} in Appendix \ref{sec:appendix-flare}). This rebrightening is not an expectation of the standard fireball model, and is likely due to an external reverse shock, discrete refreshed shock, or another form of energy injection, possibly from a long-lived central engine. The difficulty of reproducing the rebrightening lies in the rapid rise steeper than $t^3$ and the short variability timescale $\Delta T/T_\textrm{peak}$\,$\approx$\,$0.3$ of the rise.

\begin{figure}
    \centering
\includegraphics[width=\columnwidth]{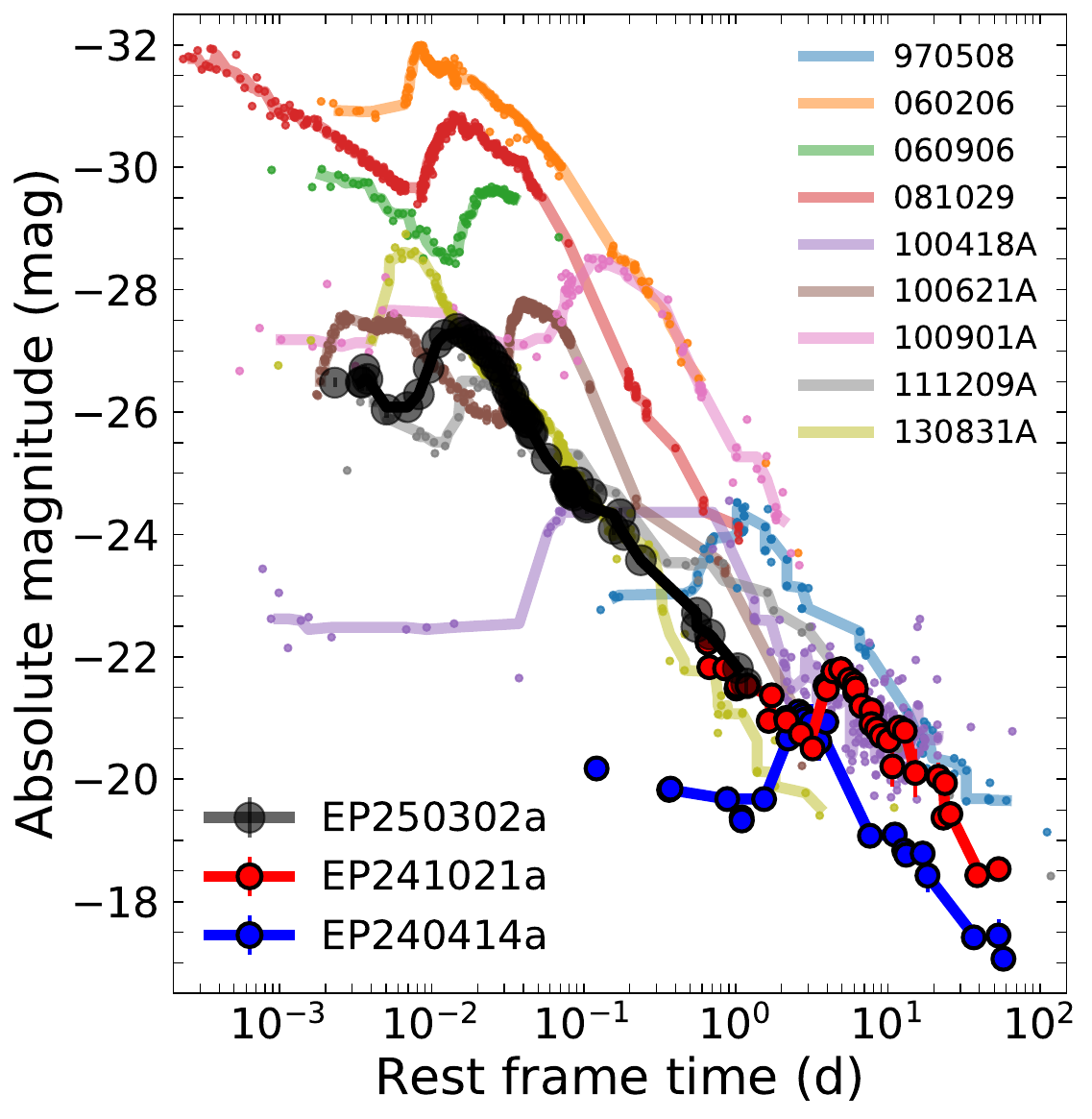}
    \caption{
    Rest-frame optical lightcurves ($r$-band) of gamma-ray bursts exhibiting a rapid and significant rebrightening. These are compared to EP240414a \citep{srivastav2024,vandalen2024}, EP241021a \citep{Busmann2025}, and EP250302a (This work and \citealt{Fu2026}). The sample shown includes: GRBs 970508 \citep{Pian1998},  060206 \citep{Wozniak2006}, 060906 \citep{Cenko2009dark}, 081029 \citep{Nardini2011}, 100418A \citep{Marshall2011,deUgartePostigo2018}, 100621A \citep{Greiner2013}, 100901A \citep{Gorbovskoy2012}, 111209A \citep{Kann2018}, and 130831A \citep{Cano2014}. The solid lines show the data smoothed using a median filter. 
    Reproduced from \citet{deUgartePostigo2018} and \citet{Busmann2025}.}
    \label{fig:rebrighten}
\end{figure}

However, such a rebrightening is not completely unprecedented. Similarly steep optical rebrightenings of $\Delta m$\,$\gtrsim$\,$1$ mag were identified in a few instances by \citet{deUgartePostigo2018}. In Figure \ref{fig:rebrighten}, we display the rest frame optical lightcurves of GRBs exhibiting a steep optical rebrightening. The data is compiled from \citet{Kann2006,Kann2010,Kann2011}\footnote{\url{https://github.com/steveschulze/kann_optical_afterglows}} and \citet{Dainotti2024}\footnote{\url{https://grblc-catalog.streamlit.app/}}. In addition to their steep rises, these events also show clear periods of shallow decay at peak, followed by a steeper decay after the end of the shallow phase \citep[see, e.g.,][]{Greiner2013}. EP250302a clearly displays a similar behavior to these GRBs both in terms of the rise timescale, rebrightening significance, shallow decay at peak, and duration of the rebrightening.

In addition, steep and significant optical rebrightenings have been observed in a handful of EP transients with various explanations invoked in the literature. These events include EP240414a at $z$\,$=$\,$0.4$ \citep{vandalen2024,srivastav2024,Sun2024} and EP241021a at $z$\,$=$\,$0.75$ \citep{Busmann2025,Yadav2025,Gianfagna2025,Shu2025,Wu2025}. EP250302a ($z$\,$=$\,$1.131$) is another good example of an EP transient showing this rebrightening feature, which occurs at significantly earlier times with better agreement to the timescale observed from GRBs (Figure \ref{fig:rebrighten}). Further observations of EP transients at $<$\,$1$ d are required to determine whether they show a higher fraction of this rebrightening behavior when compared to the more standard GRB population.

These other EP transients (EP240414a and EP241021a) are generally interpreted as non-standard GRBs, potentially displaying either choked or off-axis jets \citep{Zheng2025,Hamidani2025,Gianfagna2025} or having significant trailing ejecta that produces a refreshed shock \citep{srivastav2024,Busmann2025}. 
In general, there is an overall agreement that they are produced by similar massive star progenitors to GRBs \citep[e.g.,][]{OConnor2025EP-z}. This is supported by the detection of Type Ic-BL supernova following a handful of EP transients \citep{vandalen2024,Rastinejad2025EP,Eyles-Ferris2025EP,Li2025,EP250304a-SN-GCN,Srinivasaragavan2025EP0108a,Srinivasaragavan2026,Quirola2026sn,vanHoof2026,Cotter2026,OConnor2026,Yuan2026,Rastinejad2026,MartinCarrillo2026,Chen2026}. As such, our interpretation of the late-time Gemini data as being due to a Ic-BL supernova (Figure \ref{fig:snmodel}) has observational support and a strong match to the expected timescale of the canonical GRB-SN 1998bw \citep{Galama1998} that is generally used as a comparison object \citep{Cano2017}.

\subsection{Constraints on the Initial Lorentz Factor}
\label{sec:decel}

The afterglow onset can be used to probe the initial Lorentz factor of the jet based on the required time for the jet to decelerate \citep[e.g.,][]{Ghirlanda2018}, which depends on the jet's initial bulk Lorentz factor $\Gamma$ and isotropic-equivalent kinetic energy $E_\textrm{kin}$ and the surrounding environment's density $n$ \citep[e.g.,][]{Sari1999}. Here we consider only a uniform density environment as favored by afterglow modeling (\S \ref{sec:afterglowfit}). Prior to the jet's deceleration, the afterglow lightcurve is sharply rising as either $t^2$ or $t^3$ depending on the synchrotron emission segment. This is in contrast to the initial shallow decay observed at $<$\,$0.017$ d ($1500$ s), see Figure \ref{eqn:plfit}. Thus, we can use the timing of the initial optical observations to constrain the jet's deceleration timescale. 

The observed (on-axis) jet deceleration time is given by \citep{Sari1999,Molinari2007,Ghisellini2010,Ghirlanda2012,Nava2013,Nappo2014}
\begin{align}
t_\textrm{dec,0} &=(1+z)\Big(\frac{17}{64\pi}\frac{E_\textrm{kin,0}}{c^5\, \Gamma_0^8\, m_\textrm{p}\, n}\Big)^{1/3}, \\
&= 260\,\Big(\frac{1+z}{2}\Big) \Gamma_{2}^{-8/3}\,E_{\textrm{kin},52}^{1/3}\,n^{-1/3} \; \rm{s}
\label{eqn:tdec}
\end{align}
where $c$ is the speed of light, $m_\textrm{p}$ is the mass of the proton, and $n$ is the density of the surrounding environment. 

The early deceleration time matches well observations of both long \citep{Ghirlanda2018} and short \citep{OConnor2020,OConnor2025} GRBs for which afterglow observations require that $\sim$\,90\% of jets decelerate before 1000 s. If we assume that the jet producing EP250302a has decelerated by $\sim$\,$460$ s after the EP trigger \citep{2025GCN.39555....1W,2025GCN.39550....1Z}, we find initial Lorentz factors in the range of $\Gamma_0$\,$\gtrsim$\,$100$ for typical assumptions for the density ($1$\,cm$^{-3}$) and kinetic energy ($10^{52}$ erg).

If we more conservatively assume the jet decelerates before $0.1$ d, when the OIR lightcurves clearly match standard afterglow closure relations (\S \ref{sec:afterglowfit}), then we instead find $\Gamma_0$\,$\gtrsim$\,$25$ for the same assumptions for the density and kinetic energy, which have an overall weak dependence on the deceleration time compared to the impact of the Lorentz factor. Even in this case, the jet producing EP250302a is ultrarelativistic. 

The above calculations assume the jet was viewed on-axis. The requirement on the Lorentz factor would only be increased if we break this assumption \citep[e.g.,][]{OConnor2025}. For an off-axis viewing angle, the observed deceleration time is delayed and increases as $( \frac{\Delta\theta}{\theta_\textrm{c}})^{8/3}$ \citep{Nakar2002}, where  $\Delta \theta$\,$=$\,$\theta_{\rm obs}$\,$-$\,$\theta_\textrm{c}$, $\theta_{\rm obs}$ is the observer's viewing angle from the core, and $\theta_\textrm{c}$ is the jet core's half-opening angle. 

This suggests the jet is both ultrarelativistic and not off-axis. This has interesting implications as it shows that ultrarelativistic on-axis bursts can produce prompt emission that does not trigger gamma-ray monitors (see Appendix \ref{sec:gammalimits}). This interpretation would support the notion that most EP transients can still be produced by ``normal'' GRB jets \citep[e.g.,][]{Busmann2025}.

\subsection{Interpretation of the Early Rebrightening}

From their earliest discovery, GRB afterglows have shown a large variety of distinct behaviors at all wavelengths \citep[e.g.,][]{Nousek2006,Stanek2007} that in some cases require extensions to the standard fireball model \citep[e.g.,][]{SariMeszaros2000,ZhangMeszaros2002,Zhang2006}. Rapid optical rebrightenings have been discovered in some of the earliest optical afterglows, such as GRB 970508 \citep{Pian1998,Galama1998rebright}. As shown in Figure \ref{fig:rebrighten}, EP250302a (Figure \ref{fig:rbandfit}) is another example of a peculiar optical afterglow rebrightening observed at early times ($\lesssim$\,$1$ d). There exist an extensive collection of theoretical scenarios invoked to explain steep afterglow rebrightening episodes, flares, and optical flashes. These episodes are not explainable by only the standard external (forward) shock emission \citep{Meszaros1997,Sari1998,Granot2002}. 

The extremely steep rise (steeper than $t^3$ with a short variability timescale $\Delta T/T_\textrm{peak}$\,$\approx$\,$0.3$) of the optical rebrightening (Figure \ref{fig:rbandfit} in Appendix \ref{sec:appendix-flare}) sets stringent constraints on the possible emission mechanisms, and we are in agreement with \citet{Fu2026} that the most natural scenario is the collision of relativistic shells, as previously proposed to explain EP240414a \citep{srivastav2024} and EP241021a \citep{Busmann2025}. A comparison to these other events is shown in Figure \ref{fig:rebrighten}. We note that, as with energy injection from the central engine, the continued refreshing of the forward shock by a power-law velocity distribution of ejecta \citep[e.g.,][]{ReesMeszaros1998,Panaitescu1998,ZhangMeszaros2002} is incapable of producing this abrupt flux enhancement \citep[see, e.g.,][]{SariMeszaros2000}. Other scenarios that are capable of producing an optical flash include the deceleration of an off-axis jet \citep[e.g.,][]{PanaitescuVestrand2008}, a two-component jet \citep{Filgas2011double,Nardini2014}, reverse shock emission \citep[e.g.,][]{Kobayashi2000,Kobayashi2003,Zhang2003}, or gravitational microlensing \citep{LoebPerna1998,GranotLoeb2001}. Some other early models that previously received significant consideration have later been shown to be incapable of producing a steep flux enhancement, such as the interaction between the fireball and a density enhancement \citep{NakarGranot2007, vanEerten2009}. We note that the rebrightening was also observed in X-rays and displays chromatic behavior which sets additional constraints that can exclude some of these other possibilities (see \citealt{Fu2026}), leaving the discrete refreshed shock the most plausible explanation. However, as an alternative, we discuss the possibility of viewing this as a reverse shock (\S \ref{sec: reverseshock}). 

\subsubsection{Alternative Scenario: Reverse Shock}
\label{sec: reverseshock}

The early rebrightening, sometimes referred to as an optical flash, in the optical afterglows of GRBs is commonly interpreted as the reverse shock (RS) emission \citep[e.g.,][]{Meszaros1997,Sari1999,Sari19998GRB990123,Kobayashi2000,Kobayashi2003,Zhang2003}. This is a natural requirement of the standard fireball model and does not require any additional energy injection. A reverse shock has been applied as the explanation for a variety of other similar optical flashes \citep[e.g.][]{Gao2015,Yi2020}. However, a reverse shock does not work in all cases and many authors favor other interpretations \citep[e.g.,][]{Marshall2011,Moin2013,Laskar2015}.

We explore whether this is a plausible mechanism to explain the optical lightcurve of EP250302a (Figure \ref{fig:rbandfit}) using the thin shell reverse shock closure relations outlined by \citet{Kobayashi2000}. We find good agreement with the slow cooling scenario outlined in Figure 3a of \citet{Kobayashi2000}, and likewise discussed by \citet{Zhang2003} and references therein. 
In this case, before the shock crossing time ($t$\,$<$\,$t_\times$), we have $\nu_\textrm{a}^\textrm{RS}$\,$<$\,$\nu_\textrm{opt}$\,$<$\,$\nu_\textrm{m}^\textrm{RS}$\,$<$\,$\nu_\textrm{c}^\textrm{RS}$, where $\nu_\textrm{opt}$ refers to the characteristic frequency of the observer frame $r$-band. 

While the shock crossing time $t_\times$ is typically considered to align with the deceleration time $t_\textrm{dec}$ of the forward shock (see \S \ref{sec:decel}), there can be a slight delay where $t_\times$\,$>$\,$t_\textrm{dec}$ for a thin shell shock. As the rebrightening is occurring at quite an early phase of the GRB, this delay may be plausible, whereas for other EP transients, such as EP241021a the significantly later rebrightening (see Figure \ref{fig:rebrighten}) cannot plausibly be explained by a standard on-axis reverse shock \citep{Busmann2025}. 

The rising phase of EP250302a can be reproduced by the pre-shock-crossing phase ($t$\,$<$\,$t_\times$). In the thin shell regime, the early rising slope is provided by $t^{3p-3/2}$, which is $t^6$ for $p$\,$=$\,$2.5$. This is in good agreement with our temporal fit $F_\nu$\,$\propto$\,$t^{-\alpha}$ to the optical lightcurve which yielded $\alpha_2$\,$=$\,$-5.8^{+1.4}_{-2.0}$ (see Appendix \ref{sec:appendix-flare}). The shock crossing time coincides with the lightcurve peak where $t_\times$\,$\approx$\,$0.024$ d. 

The ordering of the characteristic synchrotron frequencies $\nu_\textrm{a}^\textrm{RS}$\,$<$\,$\nu_\textrm{opt}$\,$<$\,$\nu_\textrm{m}^\textrm{RS}$\,$<$\,$\nu_\textrm{c}^\textrm{RS}$ is the same before and after the shock crossing time ($t$\,$>$\,$t_\times$). 
In this case, the temporal decay is independent of $p$ and decays as $\approx$\,$t^{-0.5}$, 
which is in agreement with the observed shallow decay $\alpha_3$\,$=$\,$0.53\pm0.11$) at the peak of the optical rebrightening. The transition to a steeper decay at $\sim$\,$0.05$ d occurs when $\nu_\textrm{m}^\textrm{RS}$ crosses the observed band $\nu_\textrm{opt}$. The characteristic decay is $t^{-(27p+7)/35}$, which for $p$\,$=$\,$2.5$ yields a $t^{-2.1}$ decay that also matches the observed value of $\alpha_4$\,$=$\,$2.0\pm0.09$. 

While we have found a good agreement to the temporal slopes predicted for a thin shell reverse shock, we lack spectral information before shock crossing $t$\,$<$\,$t_\times$ and before $\nu_\textrm{m}^\textrm{RS}$ crosses the observed band. However, the X-ray photon index $\Gamma_\textrm{X}$\,$=$\,$\beta_\textrm{X}+1$\,$\approx$\,$1.8$ derived during the flare is consistent with the $p$\,$\approx$\,$2.5$ scenario (see also \S \ref{sec:afterglowfit}). In this case, the value of $p$ would be the same (or at least very similar) for both the forward and reverse shocks. 

We note that after $t$\,$>$\,$t_\times$ there is almost no emission from frequencies above $\nu_\textrm{c}^\textrm{RS}$. As the X-ray lightcurve shows decaying emission in this phase (Figure \ref{fig:xray}) that is consistent with a $t^{-2.1}$ decay (\S \ref{sec:temporalfits}; \citealt{Fu2026}), it is possible that $\nu_\textrm{c}^\textrm{RS}$\,$>$\,$\nu_\textrm{X}$ until at least the transition from a reverse shock dominated to a forward shock dominated lightcurve at $\sim$\,$0.1$ d. This is further supported by the X-ray photon index and its consistency with $\nu^{(1-p)/2}$ in the regime $\nu_\textrm{m}^\textrm{RS}$\,$<$\,$\nu_\textrm{X}$\,$<$\,$\nu_\textrm{c}^\textrm{RS}$. If instead $\nu_\textrm{c}^\textrm{RS}$\,$<$\,$\nu_\textrm{X}$, the spectral index would also be expected to be steeper, i.e., $\nu^{-p/2}$\,$\approx$\,$\nu^{-1.25}$ for $p$\,$\approx$\,$2.5$. 
After $\sim$\,$0.1$ d, we no longer clearly detect any reverse shock emission and cannot provide any further constraints on the model.

\section{Conclusions}
\label{sec:conclusions}

We conducted a multi-wavelength observing campaign of EP250302a between 3.87 hr to 84 d after discovery using the Fraunhofer Telescope Wendelstein, Gemini North Telescope, \textit{Neil Gehrels Swift Observatory}, and the \textit{Chandra X-ray Observatory}. EP250302a displayed luminous prompt X-ray emission, but did not have an associated gamma-ray counterpart. Besides the lack of gamma-rays, EP250302a's multi-wavelength behavior matches that of long-duration GRBs. We find that EP250302a likely launched an on-axis ultrarelativistic outflow, potentially conflicting with the lack of gamma-rays. Our late-time Gemini data reveals an optical excess between $20$\,$-$\,$30$ d after discovery. This photometric excess is consistent with emission from a Ic-BL supernova, providing a good match to the canonical GRB-SN 1998bw. This supports growing evidence that the majority of EP/FXTs are related to massive star progenitors \citep{OConnor2025EP-z}.

\begin{acknowledgments}

B. O. thanks Ramandeep Gill, Paz Beniamini, Michael Moss, Jonathan Granot, Eleonora Troja, and Jimmy DeLaunay for useful discussions. B. O. acknowledges Dmitry Svinkin for providing the \textit{Konus-Wind} constraints on on the prompt emission. B. O. thanks Pat Slane, Vinay Kashyap, and the CXO staff, and Elena Sabbi, Jen Andrews, Julia Scharwaechter, and the Gemini Director Time Review Board for approving and scheduling the DDT requests to observe EP250302a with the \textit{Chandra X-ray Observatory} and Gemini North Telescope, respectively. N. K. acknowledges useful discussions with Alice Breeveld regarding UVOT SSS corrections.

B. O. is supported by the McWilliams Postdoctoral Fellowship in the McWilliams Center for Cosmology and Astrophysics at Carnegie Mellon University. This work benefited from travel support for M. B. from the McWilliams Visitors Program as part of the McWilliams Center for Cosmology and Astrophysics at Carnegie Mellon University. M. B. is supported by a Student Grant from the Wübben Stiftung Wissenschaft.

This work used resources on the Vera Cluster at the Pittsburgh Supercomputing Center (PSC). Vera is a dedicated cluster for the McWilliams Center for Cosmology and Astrophysics at Carnegie Mellon University. We thank the PSC staff for their support of the Vera Cluster.

This paper contains data obtained at the Wendelstein Observatory of the Ludwig-Maximilians University Munich. Funded by the Deutsche Forschungsgemeinschaft (DFG, German Research Foundation) under Germany's Excellence Strategy – EXC-2094/2 – 390783311. 
The authors acknowledge the IT Support Team of the Faculty of Physics at LMU for maintaining and supporting the HPC, storage, and computational infrastructure used to process the FTW data.

Based on observations obtained at the international Gemini Observatory, a program of NSF's OIR Lab, which is managed by the Association of Universities for Research in Astronomy (AURA) under a cooperative agreement with the National Science Foundation on behalf of the Gemini Observatory partnership: the National Science Foundation (United States), National Research Council (Canada), Agencia Nacional de Investigaci\'{o}n y Desarrollo (Chile), Ministerio de Ciencia, Tecnolog\'{i}a e Innovaci\'{o}n (Argentina), Minist\'{e}rio da Ci\^{e}ncia, Tecnologia, Inova\c{c}\~{o}es e Comunica\c{c}\~{o}es (Brazil), and Korea Astronomy and Space Science Institute (Republic of Korea). The authors wish to recognize and acknowledge the very significant cultural role and reverence that the summit of Maunakea has always had within the indigenous Hawaiian community. 

The scientific results reported in this article are based on observations made by the Chandra X-ray Observatory (CXO). This research has made use of data obtained from the Chandra Data Archive provided by the Chandra X-ray Center (CXC). This research has made use of software provided by the Chandra X-ray Center (CXC) in the application package \texttt{CIAO}. 

This work made use of data supplied by the UK \textit{Swift} Science Data Centre at the University of Leicester. 
This research has made use of the XRT Data Analysis Software (XRTDAS) developed under the responsibility of the ASI Science Data Center (ASDC), Italy. 
This research has made use of data and/or software provided by the High Energy Astrophysics Science Archive Research Center (HEASARC), which is a service of the Astrophysics Science Division at NASA/GSFC. This research has made use of the Astrophysics Data System, funded by NASA under Cooperative Agreement 80NSSC21M00561.
\end{acknowledgments}


%
\facilities{WO:2m, Gemini-North, \textit{Swift}, \textit{CXO}}

\software{\texttt{Astropy} \citep{2018AJ....156..123A,2022ApJ...935..167A}, \texttt{SExtractor} \citep{1996A&AS..117..393B},  \texttt{Photutils} \citep{Bradley2024}, \texttt{SCAMP} \citep{2006ASPC..351..112B}, \texttt{SWarp} \citep{Bertin2010},
\texttt{CIAO} \citep{Ciao}, \texttt{HEASoft}\footnote{\url{https://heasarc.gsfc.nasa.gov/docs/software/lheasoft/}}, \texttt{XSPEC} \citep{Arnaud1996}, \texttt{Dragons} \citep{Labrie2019,Labrie2023}, \texttt{SFFT} \citep{Hu2022}, \texttt{emcee} \citep{emcee}, \texttt{corner} \citep{corner}, \texttt{redback} \citep{redback}, \texttt{Bilby} \citep{bilby}, \texttt{dynesty} \citep{dynesty}, \texttt{sncosmo} \citep{sncosmo}
}


\appendix

\section{Log of Observations}

Here we report the list of observations of EP250302a analyzed as part of this work. The observations are tabulated in Tables~\ref{tab: observationsPhot} and \ref{tab: observationsXray}. 

\clearpage
\startlongtable
\begin{deluxetable*}{lccccccc}
\tablewidth{0pt}
\tablecaption{Log of optical and near-infrared observations of EP250302a recorded from the EP/WXT trigger as $T_0+\delta T$. Photometry is reported in the AB magnitude system and is not corrected for Galactic extinction $E(B-V)$\,$=$\,$0.022$ mag \citep{Schlafly2011}. FTW and Gemini photometry is measured from \texttt{SFFT} difference images obtained using late-time templates. Image subtraction was not performed for UVOT, which, due to the host's faintness, has no impact on the reported measurements. 
Upper limits are reported at the $3\sigma$ level.
\label{tab: observationsPhot} }
\tablehead{
\colhead{\textbf{Start Time}} & \colhead{\textbf{$\delta T$} } & \colhead{\textbf{Telescope}} & \colhead{\textbf{Instrument}} &  \colhead{\textbf{Exposure}} & \colhead{\textbf{Filter}} & \colhead{\textbf{AB Magnitude}} \\ 
\colhead{\textbf{(UT)}} & \colhead{\textbf{(d)}} & \colhead{} & \colhead{} &  \colhead{\textbf{(s)}} &\colhead{}  & \colhead{\textbf{(mag)}}
}
\startdata
\hline\hline
   \multicolumn{7}{c}{\textbf{FTW}}  \\
\hline 
2025-03-02 19:28:31	& 0.1614 & FTW & 3KK & 180	 & \textit{r} & 	$19.98\pm0.03$ \\
2025-03-02 19:32:03	& 0.1638 & FTW & 3KK & 180	 & \textit{r} & 	$19.97\pm0.04$ \\
2025-03-02 19:35:35	& 0.1663 & FTW & 3KK & 180	 & \textit{r} & 	$20.03\pm0.04$ \\
2025-03-02 19:39:07	& 0.1687 & FTW & 3KK & 180	 & \textit{r} & 	$20.04\pm0.04$ \\
2025-03-02 19:42:39	& 0.1712 & FTW & 3KK & 180	 & \textit{r} & 	$20.05\pm0.04$ \\
2025-03-02 19:46:11	& 0.1736 & FTW & 3KK & 180	 & \textit{r} & 	$20.17\pm0.04$ \\
2025-03-02 19:49:43	& 0.1761 & FTW & 3KK & 180	 & \textit{r} & 	$20.14\pm0.04$ \\
2025-03-02 19:53:15	& 0.1786 & FTW & 3KK & 180	 & \textit{r} & 	$20.12\pm0.04$ \\
2025-03-02 19:56:47	& 0.1810 & FTW & 3KK & 180	 & \textit{r} & 	$20.14\pm0.04$ \\
2025-03-02 20:00:19	& 0.1835 & FTW & 3KK & 180	 & \textit{r} & 	$20.20\pm0.04$ \\
2025-03-03 03:43:57	& 0.50 & FTW & 3KK & 3060	 & \textit{r} & 	$21.26\pm0.06$	 \\
2025-03-03 19:44:21	& 1.17 & FTW & 3KK & 3600	 & \textit{r} & 	$22.35\pm0.09$ \\
2025-03-04 02:04:49	& 1.43 & FTW & 3KK & 3420	 & \textit{r} & 	$22.47\pm0.14$ \\
2025-03-04 20:18:53	& 2.19 & FTW & 3KK & 1800	 & \textit{r} & 	$23.03\pm0.20$ \\
2025-03-05 03:23:58	& 2.49 & FTW & 3KK & 2700	 & \textit{r} & 	$23.27\pm0.24$ \\
2025-03-06 04:13:16	& 3.52 & FTW & 3KK & 1620	 & \textit{r} & 	$>22.4$	 \\
2025-03-06 21:03:24	& 4.22 & FTW & 3KK & 5400	 & \textit{r} & 	$>23.5$	 \\
2025-03-08 01:12:57	& 5.40 & FTW & 3KK & 10440 & \textit{r} &     $>23.9$	 \\
2025-03-02 19:28:31	& 0.1614 & FTW & 3KK & 180	 & \textit{i} & 	$19.74\pm0.03$ \\
2025-03-02 19:32:03	& 0.1638 & FTW & 3KK & 180	 & \textit{i} & 	$19.73\pm0.04$ \\
2025-03-02 19:35:35	& 0.1663 & FTW & 3KK & 180	 & \textit{i} & 	$19.87\pm0.04$ \\
2025-03-02 19:39:07	& 0.1687 & FTW & 3KK & 180	 & \textit{i} & 	$19.89\pm0.04$ \\
2025-03-02 19:42:39	& 0.1712 & FTW & 3KK & 180	 & \textit{i} & 	$19.89\pm0.05$ \\
2025-03-02 19:46:11	& 0.1736 & FTW & 3KK & 180	 & \textit{i} & 	$19.94\pm0.05$ \\
2025-03-02 19:49:43	& 0.1761 & FTW & 3KK & 80	 & \textit{i} & 	$19.88\pm0.05$ \\
2025-03-02 19:53:15	& 0.1786 & FTW & 3KK & 180	 & \textit{i} & 	$19.91\pm0.05$ \\
2025-03-02 19:56:47	& 0.1810 & FTW & 3KK & 180	 & \textit{i} & 	$20.00\pm0.05$ \\
2025-03-02 20:00:19	& 0.1835 & FTW & 3KK & 180	 & \textit{i} & 	$19.90\pm 0.05$ \\
2025-03-03 03:43:57	& 0.50 & FTW & 3KK & 3060	 & \textit{i} & 	$21.08\pm 0.07$ \\
2025-03-03 19:44:21	& 1.17 & FTW & 3KK & 3600	 & \textit{i} & 	$22.13\pm 0.09$ \\
2025-03-04 02:04:49	& 1.43 & FTW & 3KK & 3420	 & \textit{i} &     $22.34\pm 0.17$ \\
2025-03-04 20:18:53	& 2.19 & FTW & 3KK & 1800	 & \textit{i} & 	$23.00\pm 0.20$ \\
2025-03-05 03:23:58	& 2.49 & FTW & 3KK & 2700	 & \textit{i} & 	$23.24\pm 0.30$ \\
2025-03-06 04:13:16	& 3.52 & FTW & 3KK & 2160	 & \textit{i} & 	$>22.0$	 \\
2025-03-06 21:03:24	& 4.22 & FTW & 3KK & 5400	 & \textit{i} & 	$>23.2$	 \\
2025-03-08 01:12:57	& 5.40 & FTW & 3KK & 10260 & \textit{i} & 	$>23.6$	 \\
2025-04-30 05:32:09 & 58.58 & FTW & 3KK & 41040 & \textit{i} & $>25.1$ \\
2025-03-02 19:28:45	& 0.1616 & FTW & 3KK & 169	 & \textit{J} & 	$19.36\pm0.08$ \\
2025-03-02 19:32:17	& 0.1640 & FTW & 3KK & 169	 & \textit{J} & 	$19.47\pm0.10$ \\
2025-03-02 19:35:49	& 0.1664 & FTW & 3KK & 169	 & \textit{J} & 	$19.30\pm0.08$ \\
2025-03-02 19:39:21	& 0.1689 & FTW & 3KK & 169	 & \textit{J} & 	$19.35\pm0.09$ \\
2025-03-02 19:42:53	& 0.1714 & FTW & 3KK & 169	 & \textit{J} & 	$19.34\pm0.09$ \\
2025-03-02 19:46:25	& 0.1738 & FTW & 3KK & 169	 & \textit{J} & 	$19.50\pm0.11$ \\
2025-03-02 19:49:57	& 0.1763 & FTW & 3KK & 169	 & \textit{J} & 	$19.42\pm0.10$ \\
2025-03-02 19:53:29	& 0.1788 & FTW & 3KK & 169	 & \textit{J} & 	$19.29\pm0.10$ \\
2025-03-02 19:57:01	& 0.1812 & FTW & 3KK & 169	 & \textit{J} & 	$19.53\pm0.13$ \\
2025-03-02 20:00:33	& 0.1837 & FTW & 3KK & 169	 & \textit{J} & 	$19.57\pm0.12$ \\
2025-03-03 03:44:11	& 0.50 & FTW & 3KK & 2885	 & \textit{J} & 	$20.61\pm0.15$ \\
2025-03-03 19:44:35	& 1.17 & FTW & 3KK & 3394	 & \textit{J} & 	$21.8\pm0.3$ \\
2025-03-04 02:05:03	& 1.43 & FTW & 3KK & 3055	 & \textit{J} & 	$>21.7$	 \\
2025-03-04 20:19:06	& 2.19 & FTW & 3KK & 1697	 & \textit{J} & 	$>22.1$	 \\
2025-03-05 03:24:12	& 2.49 & FTW & 3KK & 2546	 & \textit{J} & 	$>22.1$	 \\
2025-03-06 04:13:30	& 3.52 & FTW & 3KK & 2683	 & \textit{J} & 	$>21.1$	 \\
2025-03-06 21:03:39	& 4.22 & FTW & 3KK & 5092	 & \textit{J} & 	$>21.9$	 \\
2025-03-08 01:13:11	& 5.40 & FTW & 3KK & 9844	 & \textit{J} & 	$>22.5$  \\
\hline \hline
    \multicolumn{7}{c}{\textbf{Gemini}}  \\
\hline
 2025-03-11 09:00:57 & 8.72 & Gemini & GMOS-N & 2700 &  \textit{i} & $25.0\pm0.10$ \\
 2025-03-22 10:21:23 & 19.78 & Gemini & GMOS-N & 2250 &  \textit{i} & $25.70\pm0.10$ \\
 2025-03-30 09:31:55 & 27.75 & Gemini & GMOS-N & 2250 &  \textit{i} & $25.63\pm0.10$ \\
 2025-05-25 07:42:23 & 83.67$^a$ & Gemini & GMOS-N & 2250 &  \textit{i} & ... \\
\hline \hline
    \multicolumn{7}{c}{\textbf{UVOT}}  \\
\hline
2025-03-02 16:36:55 & 0.0422 & \textit{Swift} & UVOT & 631 & \textit{u} &  $18.66\pm0.05$ \\
2025-03-02 17:48:16 & 0.0918 & \textit{Swift} & UVOT & 398 & \textit{u} &  $19.81\pm0.13$ \\
2025-03-04 00:59:56 & 1.39$^b$ & \textit{Swift} & UVOT & 1689 & \textit{u} & --  \\
2025-03-04 02:35:09 & 1.46$^b$ & \textit{Swift} & UVOT & 547 & \textit{u} &  -- \\
2025-03-04 07:39:11 &  1.67 & \textit{Swift} & UVOT & 876 & \textit{u} &  $>21.5$ \\
2025-03-04 10:24:29 & 1.78$^b$ & \textit{Swift} & UVOT & 271 & \textit{u} &  -- \\
2025-03-05 12:59:54 & 2.89$^b$ & \textit{Swift} & UVOT & 656 & \textit{u} &  -- \\
2025-03-06 07:55:22  & 3.68$^b$ & \textit{Swift} & UVOT & 1397 & \textit{u} & --  \\
2025-03-06 09:27:50 & 3.74 & \textit{Swift} & UVOT & 366 & \textit{u} &  $>21.0$ \\
2025-03-06 17:24:26 & 4.08$^b$ & \textit{Swift} & UVOT & 860 & \textit{u} & --  \\
2025-03-06 20:23:39 & 4.20 & \textit{Swift} & UVOT & 1675 & \textit{u} &  $>21.9$ \\
2025-03-09 12:24:21 & 6.87$^b$ & \textit{Swift} & UVOT & 433 & \textit{u} & --  \\
2025-03-09 15:31:33 & 7.00$^b$ & \textit{Swift} & UVOT & 466 & \textit{u} & --  \\
2025-03-09 20:13:27 & 7.19$^b$ & \textit{Swift} & UVOT & 564 & \textit{u} & --  \\
2025-03-12 06:26:19 & 9.62$^b$ & \textit{Swift} & UVOT & 454 & \textit{u} &--\\
2025-03-12 17:01:00 & 10.06 & \textit{Swift} & UVOT & 1595 & \textit{u} &  $>21.8$ \\
2025-03-12 20:17:37 & 10.20$^b$ & \textit{Swift} & UVOT & 378 & \textit{u} & --  \\
\hline \hline
\enddata
\tablecomments{
$^a$Template epoch for image subtraction. 
$^b$UVOT snapshots impacted by SSS patches \citep{Breeveld2010}, see \S \ref{sec:uvot}. 
}
\end{deluxetable*}

\begin{table*}[ht]
\centering
\caption{Log of X-ray observations of EP250302a used in this work. 
}
\label{tab: observationsXray}
\begin{tabular}{lcccccc}
\hline\hline
\textbf{Start Time (UT)} & \textbf{$\delta T$ (d)} & \textbf{Telescope} & \textbf{Instrument} &  \textbf{Exposure (ks)} & \textbf{ObsID} \\ 
\hline
2025-03-02 16:35:05 & 0.04 & \textit{Swift} & XRT  & 1.06  & 1958001  \\
2025-03-04 00:57:02 & 1.38 & \textit{Swift} & XRT  & 2.28 & 1958002  \\
2025-03-04 07:37:02 & 1.67 & \textit{Swift} & XRT  & 1.18 & 1958003  \\
2025-03-05 08:32:00 & 2.70 & \textit{Swift} & XRT  & 0.67 & 1958004  \\
2025-03-06 07:51:00 & 3.68 & \textit{Swift} & XRT  & 4.39 & 1958005  \\
2025-03-09 12:21:00 & 6.86 & \textit{Swift} & XRT  & 1.50 & 1958006  \\
2025-03-12 06:25:00 & 9.62 & \textit{Swift} & XRT  & 2.48 & 1958007 \\

 \hline
 2025-03-11 14:04:47 & 8.94 & \textit{Chandra} & ACIS-S & 19.82 & 30840 \\
\hline\hline
\end{tabular}
\end{table*}

\section{Gamma-ray Constraints on the Prompt Emission}
\label{sec:gammalimits}

The soft X-ray transient EP250302a did not trigger any other high-energy monitors. Unfortunately, \textit{Swift} was in the South Atlantic Anomaly (SAA) and could not acquire any data, while the source was Earth occulted for \textit{Fermi}. However, \textit{Konus-Wind} was fortunately able to provide constraints on the prompt gamma-ray properties. \textit{Konus-Wind} was observing the entire sky and set a 90\% confidence upper limit of $<$\,$1.8\times10^{-7}$ erg cm$^{-2}$ s$^{-1}$ to the peak flux in the $10$\,$-$\,$1,000$ keV band using 2.944 s time bins (Dmitry Svinkin, private communication). Here we have assumed a Band function \citep{Band1993} spectral shape with peak energy $E_\textrm{p}$\,$=$\,$300$ keV and low-energy and high-energy photon indices $-1$ and $-2.5$, respectively \citep[e.g.,][]{Nava2011}. The limit on the isotropic-equivalent gamma-ray energy is $E_\textrm{iso}$\,$\lesssim$\,$2.4\times10^{51}$ erg (90\% CL) in the rest frame $1$\,$-$\,$10,000$ keV energy range. While the 2.944 s integration time used for this limit is shorter than the 
42 s soft X-ray duration of EP250302a \citep{2025GCN.39556....1D,Fu2026}, recent joint EP-GRB detections show that the soft X-ray duration is generally longer \citep[see, e.g.,][]{Yin2024,Liu2024,Gao2024,Yin2025ep250404a}.

\begin{figure}
    \centering
\includegraphics[width=\columnwidth]{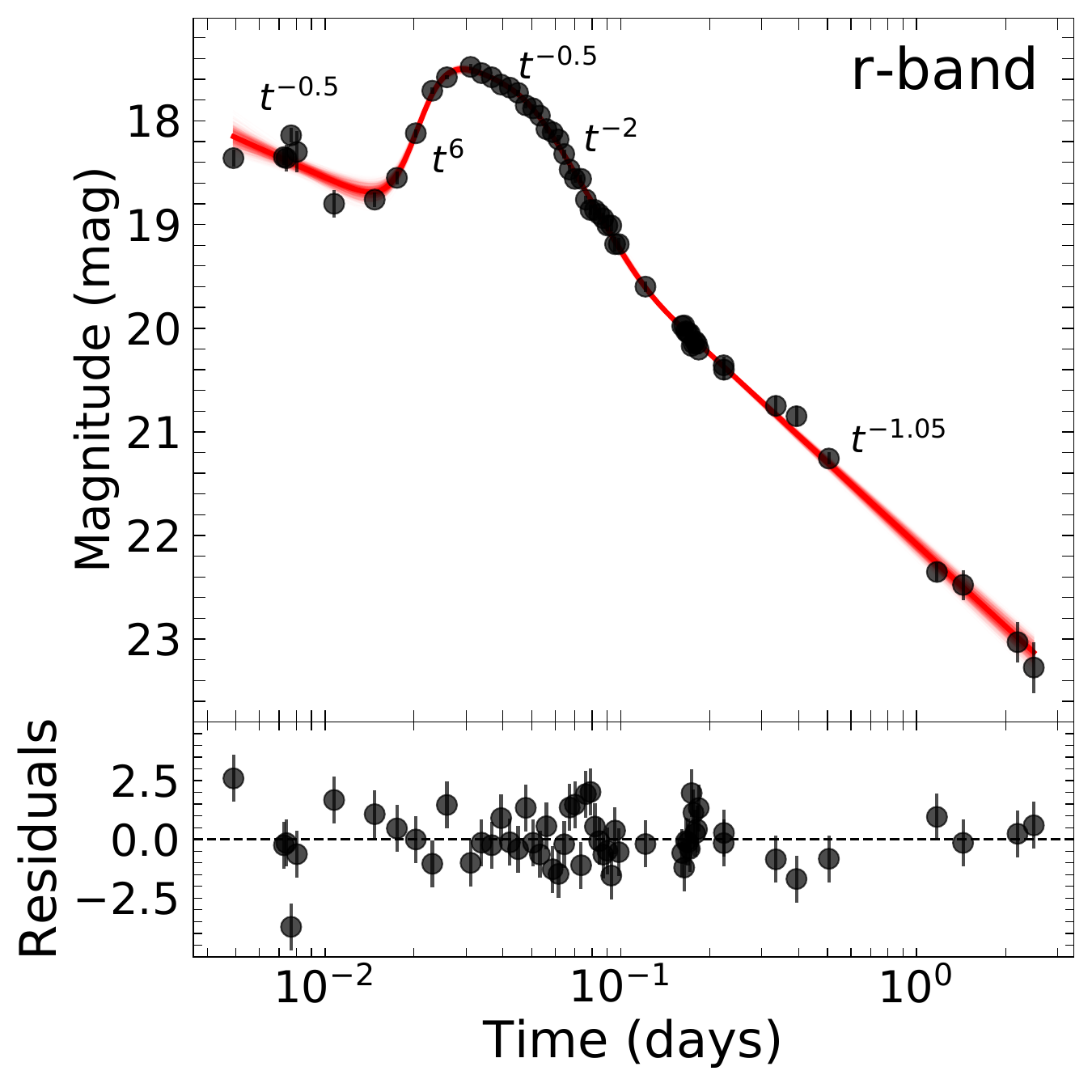}
    \caption{Temporal fit to the observed $r$-band lightcurve of EP250302a. The best fit is shown as a thick solid red line, and 1000 lightcurves randomly sampled from the posteriors are also shown in red with lower opacity. The model residuals based on the best-fit are shown as the bottom panel. The text on the plot refers to the temporal decay in flux density $F_\nu$\,$\propto$\,$t^{-\alpha}$.
    }
    \label{fig:rbandfit}
\end{figure}

\section{Early Optical Flare}
\label{sec:appendix-flare}

An early optical flare was detected by multiple teams \citep{2025GCN.39569....1K,2025GCN.39565....1P} and investigated further by  \citet{Fu2026} who interpret it as being produced by a discrete refreshed shock \citep[e.g.,][]{KumarPiran2000,Granot2003refresh,Vlasis2011}. We make use of the publicly reported rebrightening lightcurve \citep{2025GCN.39569....1K} obtained with the Nazarbayev University Transient Telescope at Assy-Turgen Astrophysical Observatory (NUTTelA-TAO; \citealt{2019JHEAp..23...14G,2022SPIE12184E..8AG}) and additional public data reported in GCN Circulars\footnote{\url{https://gcn.nasa.gov/circulars}} \citep{2025GCN.39550....1Z,2025GCN.39555....1W,2025GCN.39565....1P,2025GCN.39560....1A,2025GCN.39558....1X,2025GCN.39576....1M,2025GCN.39612....1J,2025GCN.39575....1E}. The optical data is well sampled in $r$-band starting at $\sim$\,$460$ s after the EP trigger \citep{2025GCN.39555....1W,2025GCN.39550....1Z}.

We modeled the full $r$-band lightcurve between 460 s to 3 days after the explosion with a multiply broken smoothly broken powerlaw of the form:
\begin{equation}
\label{eqn:plfit}
\! F_\nu(t)\! = \! F_0\! \left( \frac{t}{t_\textrm{b,1}} \right)^{-\alpha_1}\! \prod_{i=1}^{N}\! \left(\! 1\! +\! \left( \frac{t}{t_\textrm{b,i}} \right)^s \right)^{-(\alpha_{i+1} - \alpha_i) / s}
\end{equation}
where $F_0$ is the flux normalization, $t_\textrm{b,i}$ corresponds to the $i^\textrm{th}$ temporal break, $\alpha_{i}$ is the $i^\textrm{th}$ temporal slope, $s$ is the smoothness parameter, and $N$ is the total number of temporal breaks. The lightcurve fits were performed in flux density space using the \texttt{emcee} package \citep{emcee}. Due to the multiple segments of the lightcurve, including the early shallow phase and the rebrightening, we modeled the data with $N$\,$=$\,$4$ temporal breaks $\{t_{b,1},t_{b,2},t_{b,3},t_{b,4}\}$ which correspondingly has five decay segments $\{\alpha_1,\alpha_2,\alpha_3,\alpha_4,\alpha_5\}$. We fixed the smoothness parameter $s$\,$=$\,$10$. 

The $r$-band lightcurve exhibits an initial shallow decay ($\alpha_1$\,$=$\,$0.54\pm0.09$) starting at $\sim$\,$460$ s after the  EP trigger. A steep rise (the flare) is observed starting at $t_1$\,$=$\,$0.017\pm0.001$ d. Formally, we find a best fit value of $\alpha_2$\,$=$\,$-5.8^{+1.4}_{-2.0}$ when including no prior bounds on $\alpha_2$. This is heavily correlated (with an asymmetric posterior skewed to steeper values) with the start time of the flare with later times ($\sim$\,$0.185$ d) requiring steeper values of $\alpha_2$\,$<$\,$-6$. A earlier start to the flare of $\sim$\,$0.165$ d yields instead $\alpha_2$\,$\sim$\,$-3.5$. However, the main takeaway is that a rise steeper than $t^{3}$ is heavily required by the data. 

The flare begins to decay at $t_2$\,$=$\,$0.024\pm0.001$ d with an initially shallow decay slope ($\alpha_3$\,$=$\,$0.53\pm0.11$) around the top of the rise. This transitions to a steeper decay of $\alpha_4$\,$=$\,$2.0\pm0.09$ at $t_3$\,$=$\,$0.050\pm0.002$ d. After $t_4$\,$=$\,$0.120\pm0.006$ d the lightcurve returns to a more typical afterglow decay of $\alpha_5$\,$=$\,$1.05\pm0.04$. 
The best-fit has $\chi^2$\,$=$\,$69$ for 46 degrees of freedom (\textrm{dof}) yielding a reduced chi-squared of $\chi^2/\textrm{dof}$\,$=$\,$1.5$. The lightcurve fit and its residuals are shown in Figure \ref{fig:rbandfit}.

\section{Supernova Modeling Results}
\label{sec:appendix-sn-model-corner}

\begin{figure*}
    \centering
    \includegraphics[width=0.45\linewidth]{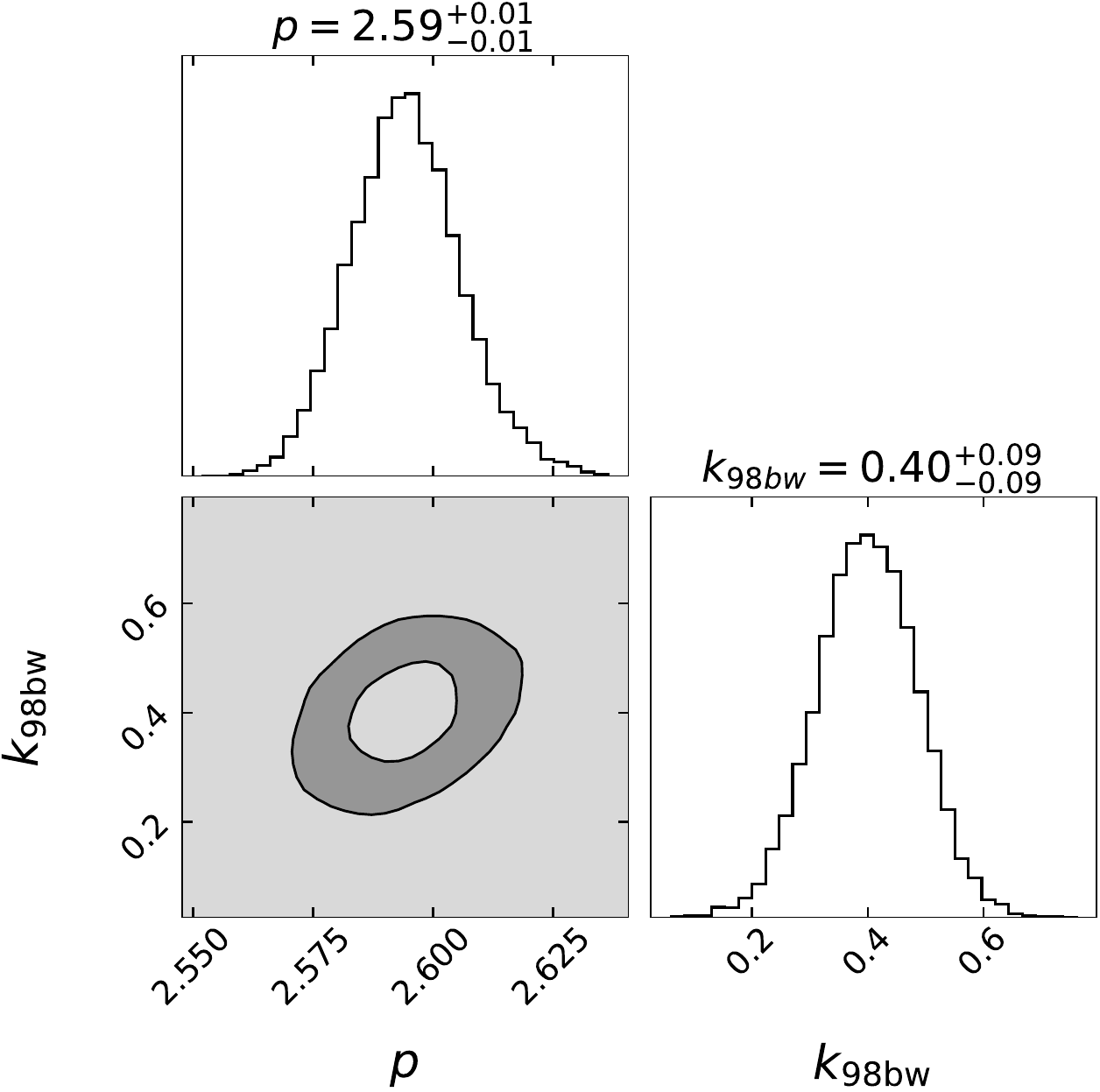}
    \includegraphics[width=0.45\linewidth]{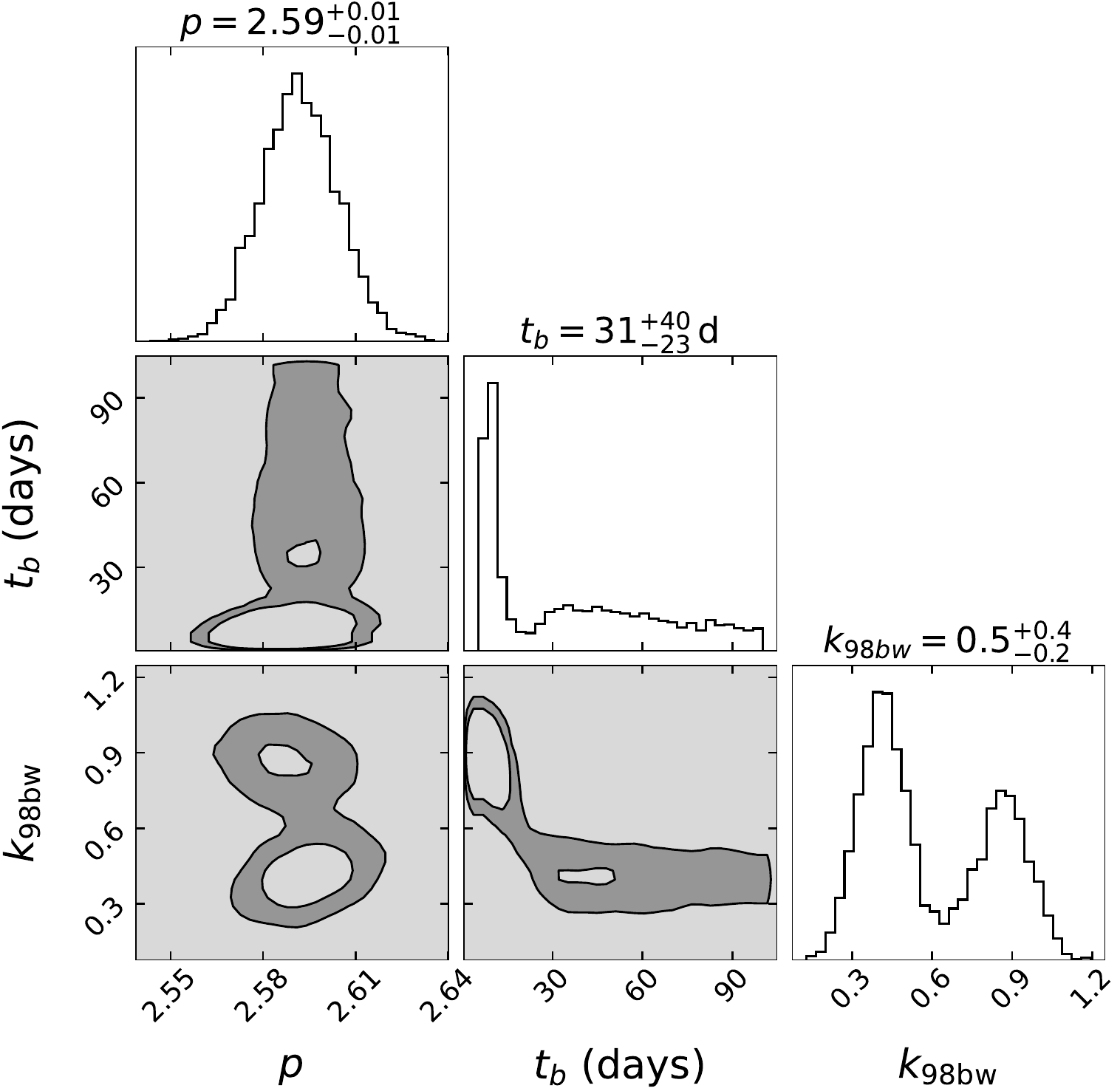}
    \caption{Corner plots showing the results of our lightcurve modeling using the combination of an afterglow plus SN 1998bw. The left corner plot is for a single powerlaw afterglow component, while the right corner plot allows for a jet-break (see \S \ref{sec:afterglowfit} for details).}
    \label{fig:corner-sn}
\end{figure*}

In Figure \ref{fig:corner-sn}, we display the resulting corner plot from our modeling of the multi-wavelength lightcurve in \S \ref{sec:snmodel}.

\bibliography{bib}{}
\bibliographystyle{aasjournal}



\end{document}